\documentclass[aps,prd,preprint,draft,showpacs]{revtex4}
\usepackage{graphicx,epsf,amssymb}
\newcommand{\be}{\begin{equation}} 
\newcommand{\ee}{\end{equation}}

\newcommand{\bea}{\begin{eqnarray}} 
\newcommand{\eea}{\end{eqnarray}} 
\newcommand{\Tr}{{\rm Tr}}

\newcommand{\N}{{\cal{N}}}

\def\CP{\mathbb{CP}}

\newif\ifdraft
\drafttrue
\newif\ifpreprint
\preprinttrue

\def\Sect#1{Section~{\ref{#1}}}
\def\sect#1{section~{\ref{#1}}}
\def\fig#1{fig.~{\ref{#1}}}

\def\spa#1.#2{\left\langle#1\,#2\right\rangle}
\def\spb#1.#2{\left[#1\,#2\right]}

\def\Tr{\, {\rm Tr}}

\newbox\charbox
\newbox\slabox
\def\s#1{{      
        \setbox\charbox=\hbox{$#1$}
        \setbox\slabox=\hbox{$/$}
        \dimen\charbox=\ht\slabox
        \advance\dimen\charbox by -\dp\slabox
        \advance\dimen\charbox by -\ht\charbox
        \advance\dimen\charbox by \dp\charbox
        \divide\dimen\charbox by 2
        \raise-\dimen\charbox\hbox to \wd\charbox{\hss/\hss}
        \llap{$#1$}
}}

\def\eqn#1{eq.~(\ref{#1})}

\def\sign{{\mathop{\rm sign}\nolimits}}

\def\mod{\mathop{\rm mod}\nolimits}

\def\sandp#1.#2.#3{%
\left\langle\smash{#1}{\vphantom1}^{+}\right|{#2}%
\left|\smash{#3}{\vphantom1}^{+}\right\rangle}
\def\ksl{\s{k}}

\newbox\ourfigbox
\def\SizedFigureWithCaption#1#2#3{%
\setbox\ourfigbox
  \hbox{\hss\epsfxsize #1 \epsfbox{#2}\hss}
\hbox to \wd\ourfigbox{\vbox{\noindent\copy\ourfigbox\break
\vskip -6mm      \hbox to \wd\ourfigbox{\hss#3\hss}}}
}
\def\SizedFigure#1#2{%
\null\epsfxsize #1 \epsfbox{#2}\hss
}

\begin{document}

\ifpreprint
hep-th/0406133
\hfill UCLA/04/TEP/26
\hfill Saclay/SPhT--T04/079
\hfill  NSF-KITP-04-74
\fi

\title{Twistor-Space Recursive Formulation of Gauge-Theory Amplitudes}

\author{Iosif Bena and Zvi Bern} 
\affiliation{Department of Physics and Astronomy, UCLA, Los Angeles, CA
90095--1547\\
{\tt iosif, bern@physics.ucla.edu}}

\author{David A. Kosower} 
\affiliation{Service de Physique, CEA--Saclay, 
          F--91191 Gif-sur-Yvette cedex, France\\
{\tt kosower@spht.saclay.cea.fr}}

\date{June 15, 2004}

\begin{abstract}
Using twistor space intuition, Cachazo, Svrcek and Witten 
presented novel diagrammatic rules for gauge-theory amplitudes,
expressed in terms of maximally helicity-violating (MHV)
vertices.  We define non-MHV vertices, and show how to use them
to give a recursive
construction of these amplitudes. 
 We also use them to illustrate the equivalence of
various twistor-space prescriptions, and to determine the associated
combinatoric factors.
\end{abstract}

\pacs{11.15.Bt, 11.25.Db, 11.25.Tq, 11.55.Bq, 12.38.Bx \hspace{1cm}}

\maketitle



\renewcommand{\thefootnote}{\arabic{footnote}}
\setcounter{footnote}{0}

\section{Introduction}
\label{IntroSection}

\vskip -.4  cm

In a striking recent paper~\cite{WittenTopologicalString}, Witten
suggested that tree-level
amplitudes of non-Abelian gauge theories can be obtained
 by integrating over the moduli space
of certain $D$-instantons in the open-string topological B-model on
(super) twistor space 
$\CP^{3|4}$. This proposal generalizes
Nair's earlier construction~\cite{Nair} of maximally
helicity-violating amplitudes.
It is a very promising step towards
understanding why amplitudes in unbroken gauge theories are so much
simpler than suggested by individual Feynman diagrams.

The best-known example of amplitudes with a remarkably simple
structure are the Parke--Taylor amplitudes~\cite{ParkeTaylor} 
of QCD.  These are
amplitudes with two negative-helicity gluons and an arbitrary number
of positive-helicity ones (or their parity conjugates), the so-called
maximally helicity-violating (MHV) amplitudes.  Inspired by the
twistor-space structure underlying gauge-theory amplitudes in Witten's
proposal, Cachazo, Svrcek, and Witten (CSW)~\cite{CSW} have proposed a
novel way to use off-shell continuations of MHV amplitudes to
construct arbitrary tree-level amplitudes. This construction makes
manifest the factorization on multi-particle poles.  Although no-one
has yet given a direct derivation of the CSW construction from
a Lagrangian, the combination of the correct pole
structure --- that is, tree-level unitarity --- and Lorentz invariance
leaves little doubt that it is correct.  It is worth noting that
supersymmetry does not appear to play a special role in this
construction, which is therefore valid in any gauge theory, including
QCD~\cite{Khoze}.

\def\MHVbar{$\overline{\rm MHV}$} In the original twistor-space
formulation~\cite{WittenTopologicalString},
one obtains tree-level amplitudes with $d+1$ negative-helicity gluons
 from D-instantons of degree $d$. These might include
both lone- and multi-instanton contributions, where the total degree
of the instantons is $d$, and disconnected instantons are joined by
twistor-space propagators.

In a series of papers Roiban, Spradlin, and Volovich~\cite{RSV1,RSV2} 
computed the
integrals over the moduli space of connected curves of degree $d$ and
found that for \MHVbar{} amplitudes ($n$-point amplitudes with $n-2$
negative helicities, that is the parity conjugates of MHV, also called
`googly-MHV') and next-to-MHV amplitudes (with three negative helicities)
this integral does reproduce
the gauge theory amplitude.  Since the integral over degree-$d$ curves
gives the full result, it appears that one can ignore disconnected
instantons.  On the other hand, the CSW construction of gauge theory
amplitudes is based on integrating over the moduli space of completely
disconnected instantons of degree one, linked by twistor space
propagators. Indeed, degree-one instantons correspond to
MHV vertices; obtaining amplitudes with an arbitrary
number of negative helicities using MHV vertices is dual in twistor space to
integrating over degree-one instantons.  We therefore seem to have
different ways of computing the amplitudes from the topological B
model.  

In order to reconcile the two different computations, Gukov, Motl, and
Neitzke~\cite{Gukov} examined the twistor space integrals over the
moduli spaces of one degree-$d$ instanton and of $d$ degree-one
instantons respectively. They argued that the former integral reduces
to one on a special locus, where the degree $d$ curve degenerates into $d$
intersecting degree one curves. Similarly, the latter integral can be argued to
reduce to one over the locus where the propagators have shrunk to zero
size, which is also the place where the $d$ degree one curves
intersect. Accordingly, it seems clear that the two prescriptions 
give the same result.

Following an analysis of degenerating twistor space curves, Gukov
et.~al.\ also suggested a set of `intermediate' prescriptions for the
amplitudes, in which one integrates over the moduli space of $k$
curves of degrees $d_i$ such that $\sum_{i=1}^{k} d_i = d$. In the
extreme cases, $k=1$ and $k=d$, this reproduces the maximally
connected and maximally disconnected calculations.  They proposed that
the amplitudes should be constructible from non-MHV vertices, though
their form was not determined.

In this paper, we construct non-MHV vertices explicitly and use them
to formulate intermediate prescriptions for tree-level
amplitudes.  Gukov et.~al.~\cite{Gukov} used properties of curves in twistor
space to investigate properties of amplitudes.  Here we employ
the connection in the opposite direction, 
using properties of amplitudes to determine the
combinations of twistor-space curves that yield valid
representations of the amplitudes.  A simple approach to finding
appropriate combinations of twistor-space curves that 
reproduce the gauge-theory amplitudes makes use of {\it skeleton
diagrams}, which are like CSW diagrams, but stripped of their external
legs.  The set of twistor-space prescriptions corresponds to the
possible reorganizations of MHV skeleton diagrams into diagrams including
non-MHV vertices as well.

We also use the non-MHV vertices to formulate the CSW construction in
a recursive fashion.  Recursive methods have proven very powerful in
QCD.  Berends and Giele~\cite{Recurrence} used recurrence relations to
give a proof of the Parke--Taylor equations~\cite{ParkeTaylor}, and
later to derive explicit expressions~\cite{WFourJet}
 for processes such as $W+4$~jet
production, the dominant background to top-quark production.  These
methods can be used both analytically and numerically.  Beyond MHV
amplitudes, they have also been used to obtain an all-$n$
form for the adjacent three-negative-helicity
amplitude~\cite{LightConeRecurrence}.  For this same class of
amplitudes, CSW presented a remarkably simple derivation
of an explicit formula for the
amplitude.  Their construction of course allows one to compute any
amplitude; but as the number of negative-helicity legs increases, the
number of diagrams and hence the computational complexity still increases
exponentially \cite{RSV2}. Just as in the older
case of Feynman diagrams, recursive methods can reduce this
computational complexity.  The recursive expressions we present in
this paper can likewise be used directly for numerical computations as 
well as analytic ones.  We will focus on amplitudes with external gluons,
but our formul\ae{} generalize in a straightforward way to amplitudes
with external fermions or scalars~\cite{Khoze}, 
in both supersymmetric and nonsupersymmetric
theories.

Alternative approaches to using string theories to compute gauge
theory amplitudes have appeared in
ref.~\cite{Berkovits,Vafa,Siegel}. Also, the twistor-space formulation
breaks manifest parity relations, such as between MHV and \MHVbar{}
amplitudes~\cite{WittenTopologicalString,RSV1,OtherGoogly}, but it has 
been shown to hold
generally for twistor-space 
amplitudes~\cite{BerkovitsMotl,RSV2,WittenParity}.

What about loop amplitudes?
Recently, Berkovits and Witten have noted~\cite{BerkovitsWitten} that
the field theory described by the topological
B-model in $\CP^{3|4}$, or alternatively by the open string theory
proposed in refs.~\cite{Berkovits,BerkovitsMotl}, is not pure $\N=4$ super-Yang
Mills theory, but rather super-Yang Mills coupled to conformal
supergravity. Although this does not affect the computation of
tree-level amplitudes (at tree level one can simply discard the
conformal supergraviton contributions to gauge-boson amplitudes), it
does affect loop amplitudes computed from twistor space, because the
conformal supergravitons can circulate in the loops.

As noted by Berkovits and Witten,
anomaly cancellation considerations constrain the possible gauge groups
in these string theories, although the remaining freedom to vary them
is not yet clear.  They considered taking the limit $k\rightarrow 0$
for the level of the current algebra in order to decouple the conformal
supergravitons.  Alternatively,
were it possible to vary the gauge
group, or perhaps the level $k$ of the current algebra, 
it should be possible to extract the leading-color contributions
at any loop order $L$ (that is, the single-trace contributions
carrying an explicit factor of $N_c^L$).   At one loop,
this would actually suffice to extract the entire amplitude, because
the subleading-color (double-trace) gauge-theory amplitudes
are algebraically determined by the
leading-color ones~\cite{Neq4Oneloop,SubleadingColorRelation}.

While it is unclear at the moment how to obtain complete higher-loop 
gauge-theory amplitudes
from a twistor-space string theory, one can approach the issue
from the field theory side.  
The unitarity-based method~\cite{UnitarityMachinery}
 builds upon tree amplitudes to produce 
results for loop amplitudes bypassing traditional Feynman-diagram
calculations.  Indeed, knowledge of the MHV tree amplitudes for
any number of external legs has enabled the computation of MHV
one-loop amplitudes for an arbitrary number of external legs, in both
$\N=4$ and $\N=1$ supersymmetric gauge 
theories~\cite{Neq4Oneloop,Neq1Oneloop}.  
Concrete knowledge of non-MHV tree amplitudes
should allow new infinite series of one-loop amplitudes to be computed,
extending our knowledge beyond the six-point amplitude.
  It would be interesting to make use
of such methods on the string-theory side to assist in
constructing a string theory providing
a dual description of pure $\N=4$ super-Yang Mills at weak coupling. The 
existence of this dual is plausible, given that the sigma model 
 dual to $\N=4$ super-Yang Mills at strong coupling exhibits signs
of integrability \cite{sigma}.

A related observation is that
the structure of one-loop amplitudes in the $\N=4$ supersymmetric
amplitudes is much simpler than one might have anticipated {\it a priori\/}.
Yet another striking result is the relation between one-loop and two-loop 
amplitudes, which effectively allows one to express the latter 
in terms of the former~\cite{IterationRelation}.

In the next section, we review the CSW construction.  In 
~\sect{SkeletonSection}, we introduce skeleton diagrams, followed
by an explicit construction of non-MHV vertices
in~\sect{NonMHVVerticesSection}.  We present a recursive form
for the amplitudes in \sect{RecurrenceSection}.  
We discuss other constructions
of amplitudes from non-MHV vertices in \sect{AlternativeSection},
and link all these constructions to twistor-space notions in 
\sect{TwistorSection}.

\section{Amplitudes from MHV Building Blocks}
\label{CSWsection}

It is convenient to write the full momentum-space tree-level amplitude 
in the $\N=4$ supersymmetric gauge theory using
a color decomposition~\cite{Color},
\begin{equation}
{\cal A}_n(\{k_i,\lambda_i,a_i\}) = 
\sum_{\sigma \in S_n/Z_n} \Tr(T^{a_{\sigma(1)}}\cdots T^{a_{\sigma(n)}})\,
A_n(\sigma(1^{\lambda_1},\ldots,n^{\lambda_n}))\,,
\label{TreeColorDecomposition}
\end{equation}
where $S_n/Z_n$ is the group of non-cyclic permutations
on $n$ symbols, and $j^{\lambda_j}$ denotes the $j$-th momentum
and helicity $\lambda_j$.  The notation $j_1+j_2$ appearing below will denote
the sum of momenta, $k_{j_1}+k_{j_2}$.
We use the normalization $\Tr(T^a T^b) = \delta^{ab}$.  The
color-ordered amplitude $A_n$ is invariant under
a cyclic permutation of its arguments.  It is the object 
which we wish to calculate
directly.

The CSW construction~\cite{CSW} builds amplitudes out of building
blocks which are off-shell continuations of the Parke--Taylor amplitudes.
\def\vo{\vphantom{1}}
These amplitudes, with two negative-helicity gluons and any number of
positive-helicity ones, are the maximally helicity-violating nonvanishing 
tree-level amplitudes
in the theory, and are therefore called MHV amplitudes.  
Using spinor products~\cite{Spinor}, we can write them in the 
the simple form,
\begin{equation}
A_n(1^+,\ldots,m_1^{-},(m_1\!+\!1)\vo^{+},\ldots,m_2^{-},
          (m_2\!+\!1)\vo^{+},\ldots,n^+) = 
  i {\spa{m_1}.{m_2}^4\over \spa1.2\spa2.3\cdots \spa{(n\!-\!1)}.n\spa{n}.1},
\end{equation}
where the two negative-helicity gluons are labeled
$m_{1,2}$. In this equation,
$\spa{i}.{j} = \spa{k_i}.{k_j}$.
We follow the standard spinor normalizations
$\spb{i}.{j} = \sign(k_i^0 k_j^0)\spa{i}.{j}^*$ and
$\spa{i}.{j}\spb{j}.{i} = 2 k_i\cdot k_j$.

The off-shell continuation of this amplitude is an MHV vertex.
The CSW prescription for the
 off-shell continuation of a momentum $k_j$ amounts to replacing
\begin{equation}
\spa{j}.{j'}  \longrightarrow \spb{q}.j \spa{j}.{j'}
 \longrightarrow \sandp{q}.{\ksl_j}.{j'},
\label{CSWOffShell}
\end{equation}
where $q$ is an arbitrary light-like reference vector, in the Parke-Taylor formula. 
(The extra factors thereby introduced will
cancel when sewing vertices to obtain an on-shell amplitude.
As shown by CSW~\cite{CSW}, on-shell amplitudes are in fact independent
of the choice of $q$.)

\def\proj{\flat}
An alternative but equivalent way of going off-shell leads
to an interpretation of the
extra momentum $q$ as the light-cone
gauge vector~\cite{NMHVpaper}.  
Observe that we can always decompose the off-shell momentum
$K$ into a sum of two massless momenta, where one is proportional to $q$,
\begin{equation}
K = k^\proj + \eta(K) q.
\end{equation}
The constraint $(k^\proj)^2 = 0$ yields
\begin{equation}
\eta(K) =  {K^2 \over 2 q\cdot K}.
\end{equation}
Of course, if $K$ goes on shell, $\eta$ vanishes.  Also, if two
off-shell vectors sum to zero, $K_1+K_2=0$, then so do the corresponding
$k^\proj$s.  This leads to the prescription for continuing MHV amplitudes
or vertices off-shell,
\begin{equation}
\spa{j}.{j'}\rightarrow \spa{\smash{j^\proj}}.{j'},
\label{OffShellPrescription}
\end{equation}
when $k_j$ is taken off shell.  It is equivalent to the CSW prescription
in the computation of on-shell amplitudes. The on-shell limit of course
just amounts to reversing the arrow.

The CSW construction replaces ordinary Feynman diagrams with diagrams
built out of MHV vertices and ordinary propagators.  Each vertex has
exactly two lines carrying negative helicity (which may be on or off
shell), and at least one line carrying positive helicity.  The propagator
takes the simple form $i/K^2$, because the physical state 
projector is effectively supplied by the
vertices.  The simplest all-gluon vertex is an amplitude with one leg taken
off shell,
\def\ks#1{\{#1\}}
\begin{equation}
A_n(1^+,\ldots,m_1^{-},(m_1\!+\!1)\vo^{+},\ldots,m_2^{-},
          (m_2\!+\!1)\vo^{+},\ldots,(n\!-\!1)\vo^+,(-K_{1,n-1})\vo^+),
\end{equation}
using the prescription~(\ref{OffShellPrescription}) for the off-shell leg.

It will be convenient to denote the projected
$k^{\proj}$ momentum built out of $-K_{1,n}$ by $\ks{1\cdots n}$, for example 
$\spa{j}.{k^\proj(-K_{1,n},q)} = \spa{j}.{\ks{1\cdots n}}$.  
The simplest vertices then have the explicit expression,
\begin{eqnarray}
&&A_n(1^+,\ldots,m_1^{-},(m_1\!+\!1)\vo^{+},\ldots,m_2^{-},
          (m_2\!+\!1)\vo^{+},\ldots,(-K_{1\cdots (n-1)})\vo^+)\ = \nonumber\\
&&\hskip 4cm  {i\spa{m_1}.{m_2}^4\over \spa1.2\spa2.3\cdots 
      \spa{(n\!-\!1)}.{\ks{1\cdots(n\!-\!1)}}\spa{\ks{1\cdots(n\!-\!1)}}.1} \,,
         \nonumber\\
&& A_n(1^+,\ldots,m_1^{-},(m_1\!+\!1)\vo^{+},\ldots,
          (-K_{1\cdots n-1})\vo^-)\ = \nonumber\\
&&\hskip 4cm 
{i\spa{m_1}.{\ks{1\cdots(n\!-\!1)}}^4 \over \spa1.2\spa2.3\cdots 
      \spa{(n\!-\!1)}.{\ks{1\cdots(n\!-\!1)}}\spa{\ks{1\cdots(n\!-\!1)}}.1} \,,
\label{Vertices}
\end{eqnarray}
where $K_{j\cdots l} = k_j+\cdots+k_l$.  We will also use the notation
$s_{j\cdots l} = K_{j\cdots l}^2$.
There are similar formul\ae{}~\cite{Khoze}, related by supersymmetry Ward 
identities~\cite{SWI}, for other external particle states.

\begin{figure}[t]
\centerline{\epsfxsize 3.5 truein \epsfbox{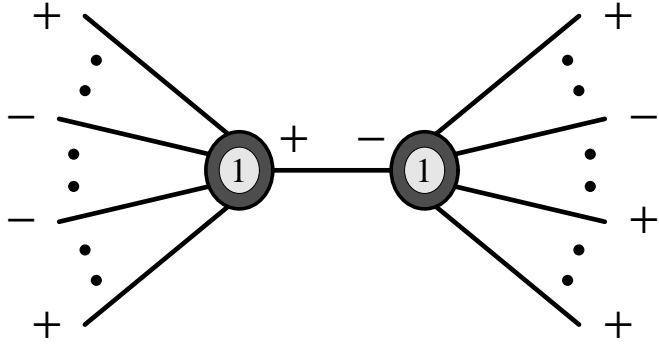}}
\caption[a]{\small A term in the CSW representation of an NMHV
amplitude.  The black dot represents the multiparticle pole
multiplying the two on-shell amplitudes.  The `1' inside the vertex
signifies that it is a basic MHV vertex corresponding to a degree 1 curve 
in twistor space.}
\label{CSWTermFigure}
\end{figure}

The CSW rules then instruct us to write down all tree diagrams with
MHV vertices, subject to the constraints that each vertex have exactly
two negative-helicity gluons and at least one positive-helicity gluon attached,
 and that each propagator connect
legs of opposite helicity.
 For amplitudes with two negative-helicity
gluons, the vertex with all legs taken on shell is then the amplitude.
For each additional negative-helicity gluon, we must add a vertex and
a propagator.  The number of vertices is thus the number of negative-helicity
gluons, less one.
For example, to compute 
amplitudes with three negative-helicity gluons, we must write down all
diagrams with two vertices.  One of the vertices has two of the external
negative-helicity gluons attached to it, while the other has only one.
An example of such a diagram is shown in 
\fig{CSWTermFigure}.

\begin{figure}[t]
\hskip 1truecm\vbox{\hbox{%
 \SizedFigure{1.8 truein}{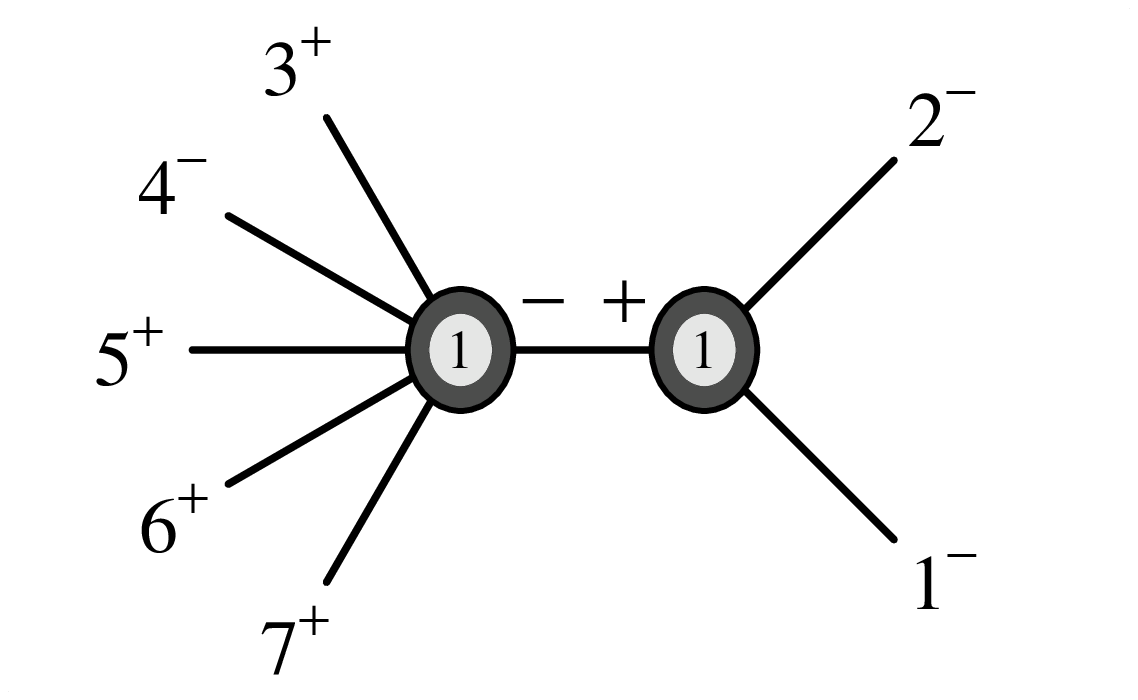}{}%
 \SizedFigure{1.8 truein}{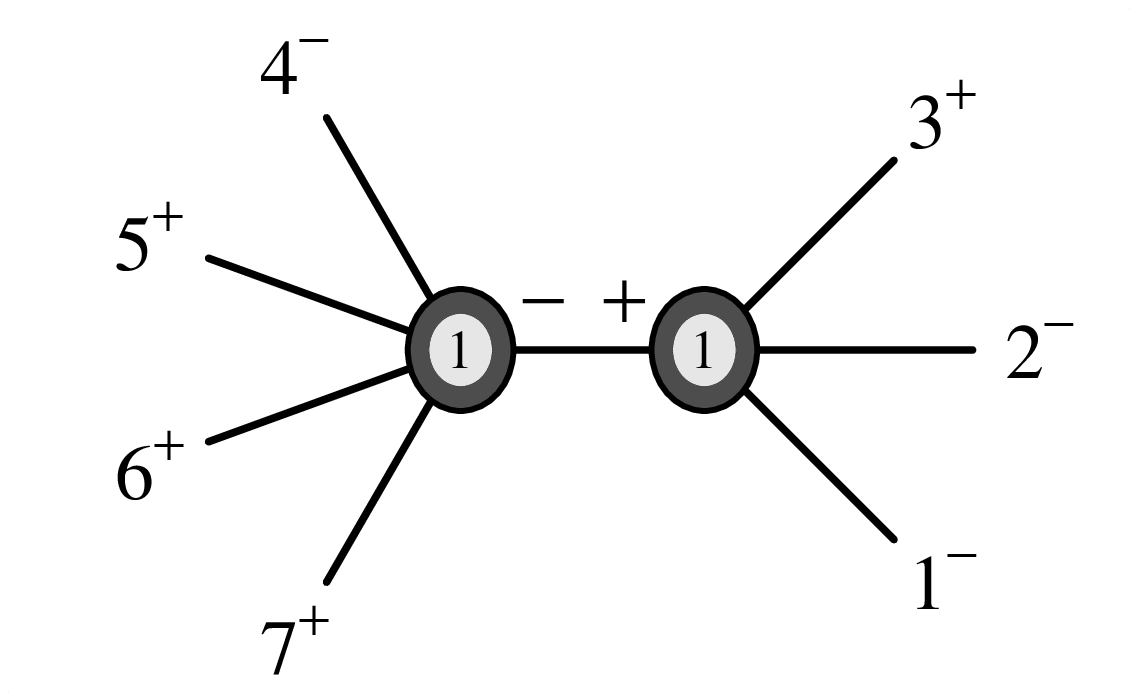}{}%
 \SizedFigure{1.8 truein}{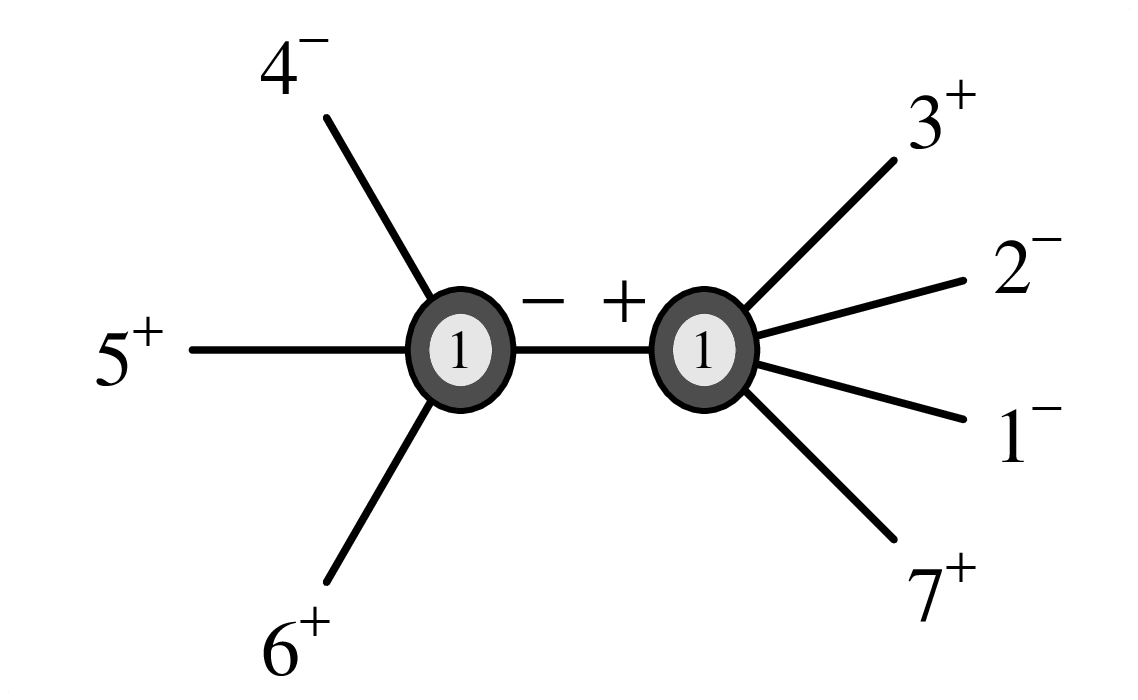}{}%
 }
\hbox{%
 \SizedFigure{1.8 truein}{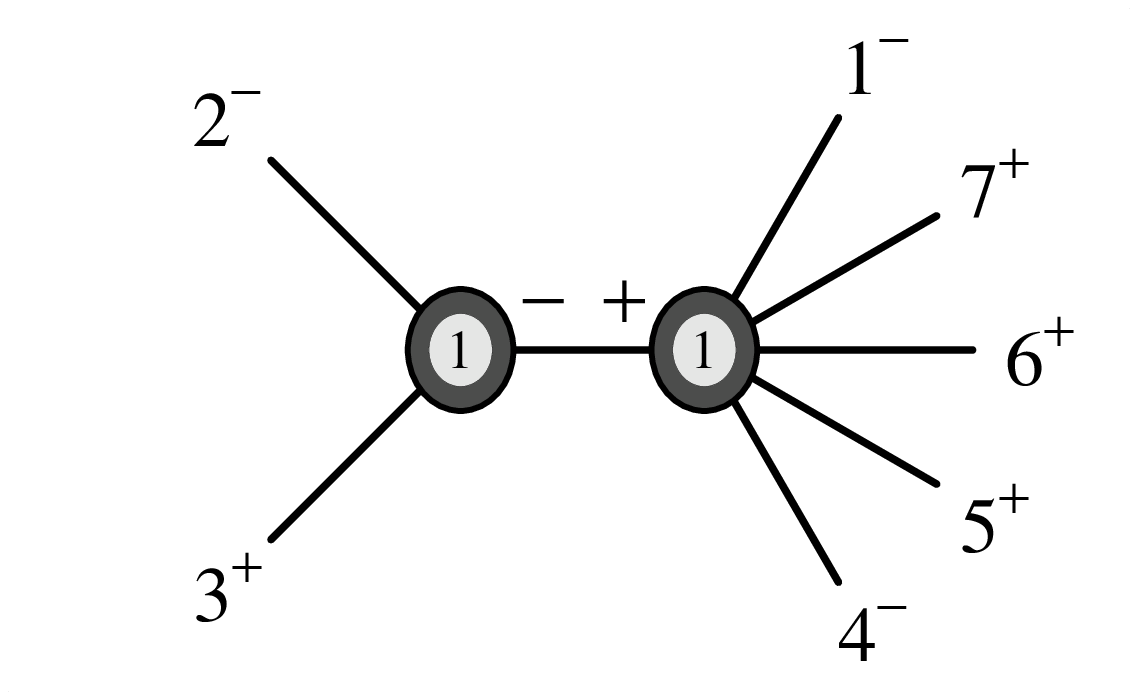}{}%
 \SizedFigure{1.8 truein}{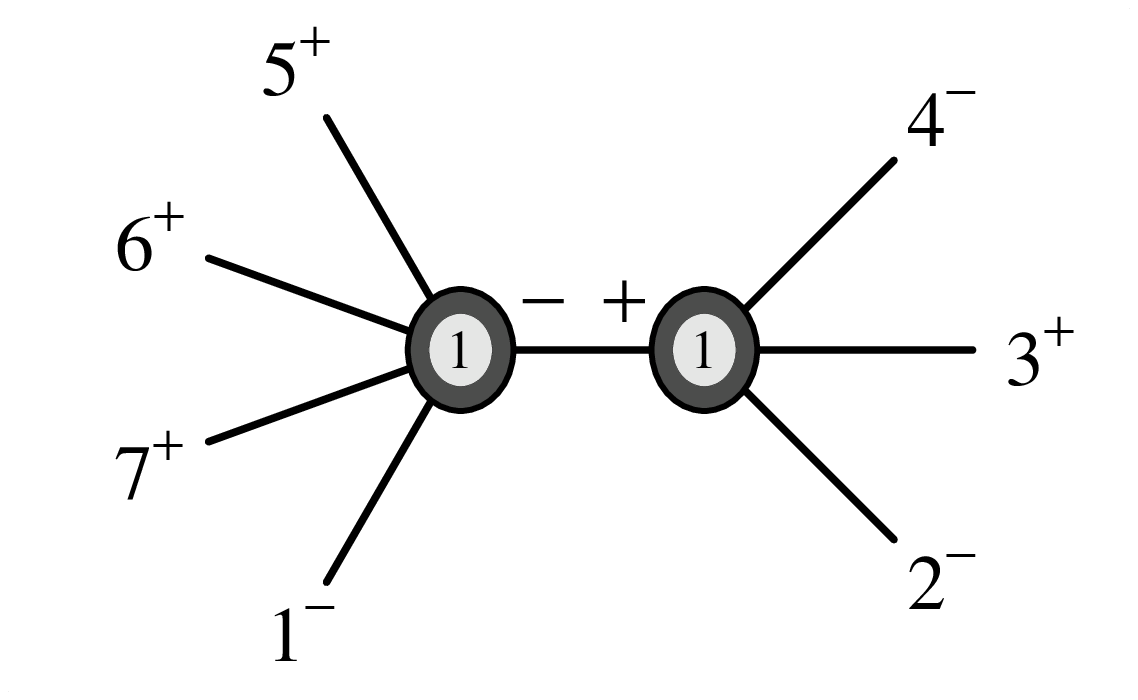}{}%
 \SizedFigure{1.8 truein}{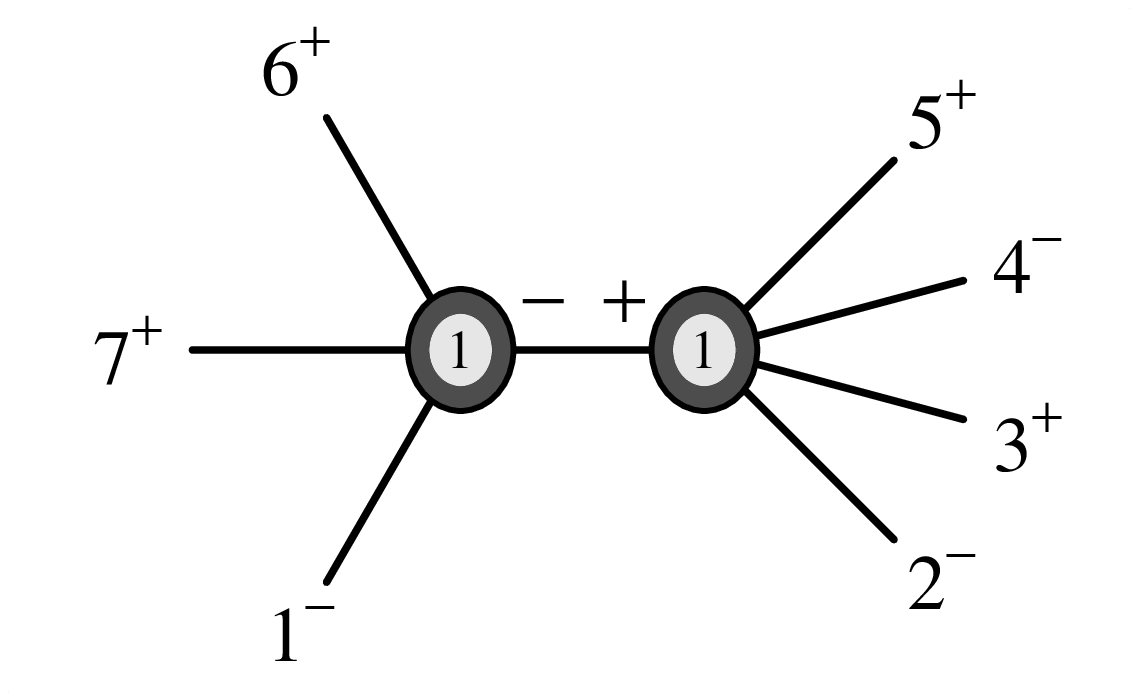}{}%
}}
\caption[a]{\small 
Examples of diagrams for the seven-point NMHV 
amplitude $A_7(1^-,2^-,3^+,4^-,5^+,6^+,7^+)$.}
\label{SevenPointFigure}
\end{figure}

\begin{figure}[b]
\hskip 1truecm\vbox{\hbox{%
 \SizedFigure{1.8 truein}{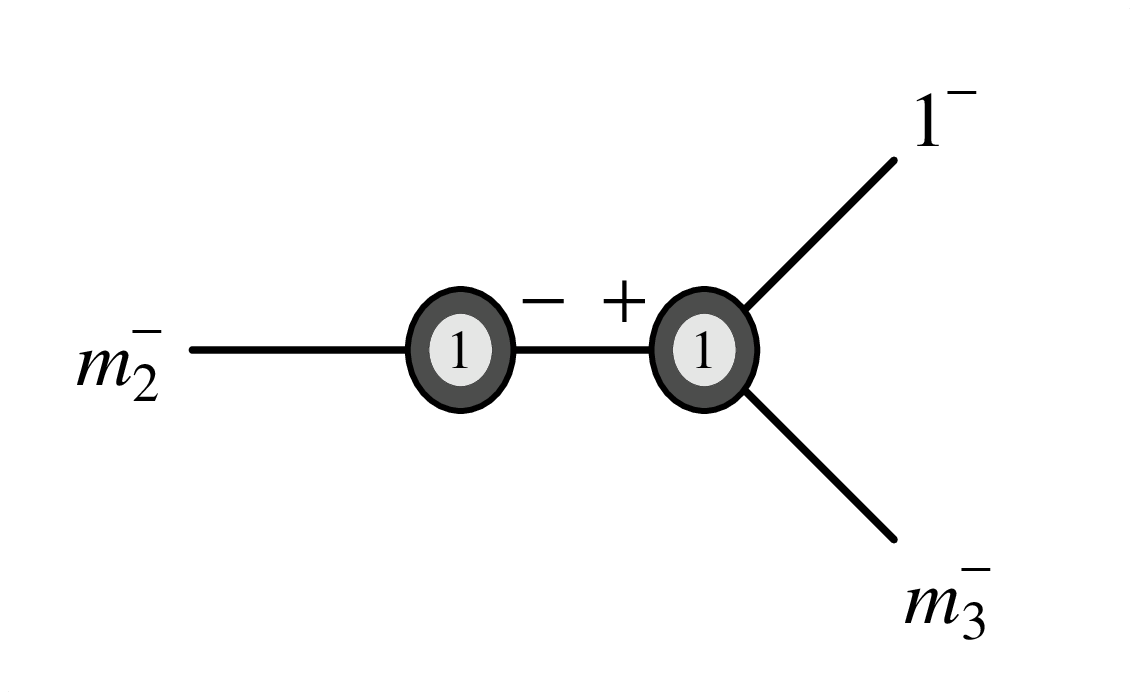}{}%
 \SizedFigure{1.8 truein}{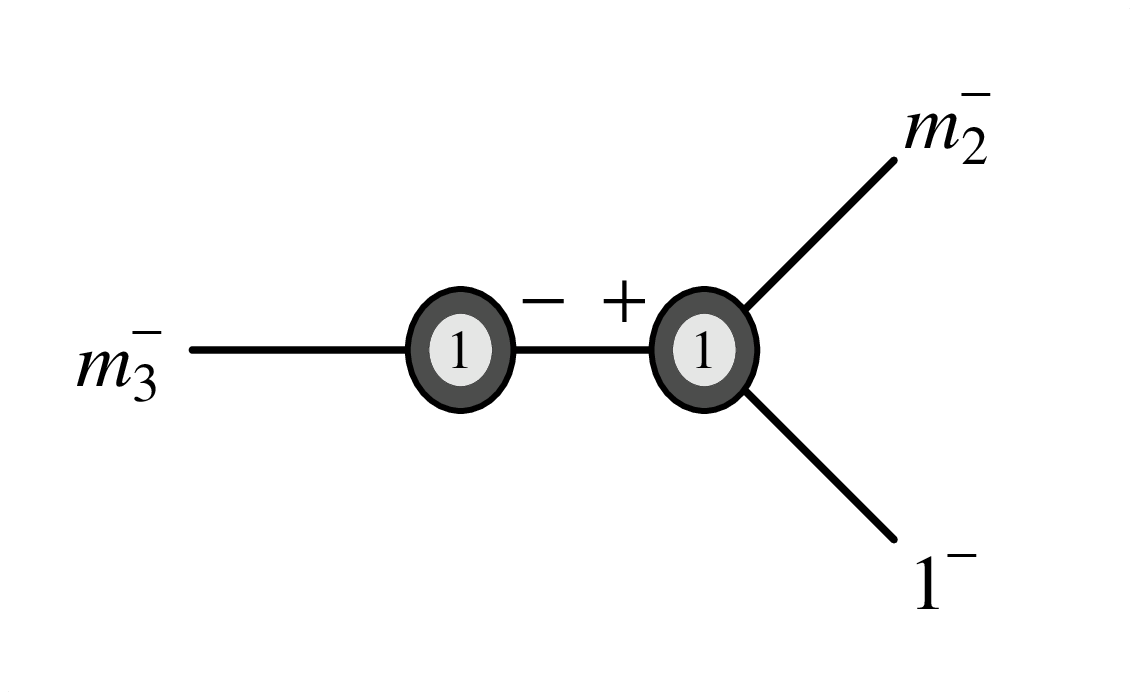}{}%
 \SizedFigure{1.8 truein}{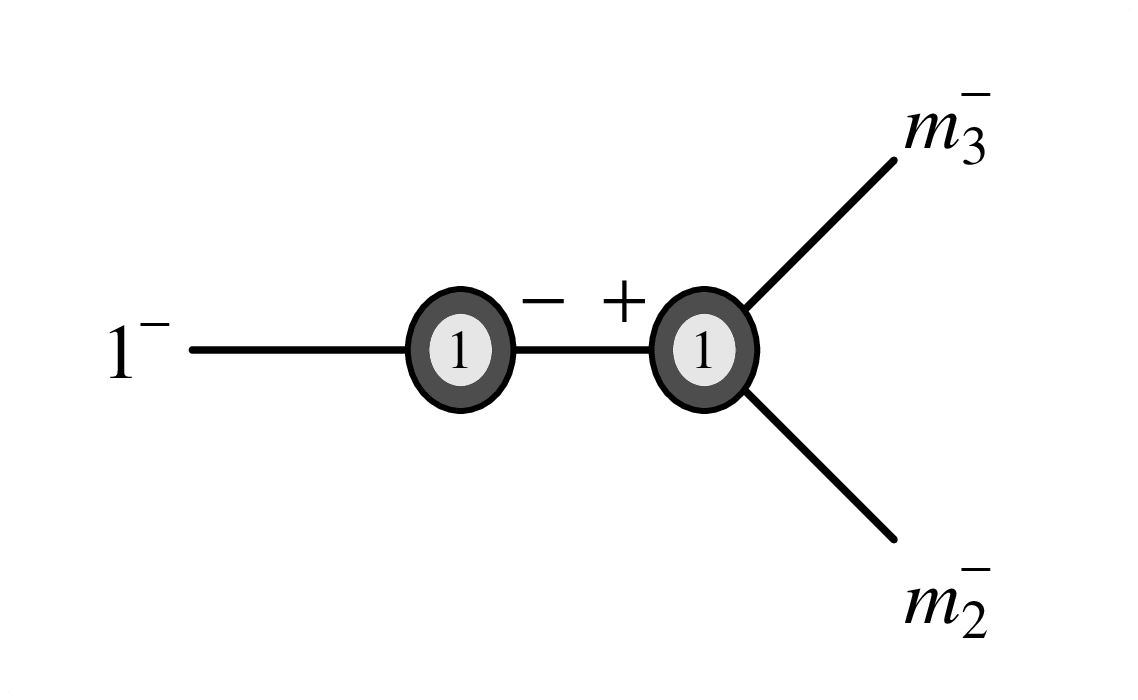}{}%
 }}
\caption[a]{\small 
The stripped diagrams for amplitudes with three negative-helicity
gluons.}
\label{ThreeMinusFigure}
\end{figure}

Some diagrams for the amplitude $A_7(1^-,2^-,3^+,4^-,5^+,6^+,7^+)$
are shown in \fig{SevenPointFigure}.  We can classify the diagrams into
three sets, according to which of the pairs of negative-helicity gluons
($(1,2)$, $(2,4)$, or $(4,1)$) are attached to the same vertex.  Indeed,
we can imagine obtaining the set of diagrams by first taking diagrams
with only the negative-helicity external gluons indicated, and then adding
the positive-helicity gluons in all possible ways consistent with the 
cyclic ordering.  Only three such diagrams are needed to generate the set of
diagrams for the seven-point amplitude, or indeed for any $n$-point
amplitude $A_n(1^-,2^+,\ldots,m_2^-,(m_2\!+\!1)\vo^+,
\ldots,m_3^-,(m_3\!+\!1)\vo^+,
\ldots,n^+)$.  The corresponding {\it stripped\/} diagrams are
shown in \fig{ThreeMinusFigure}.  Each set of stripped diagrams represents
an infinite series of amplitudes, with an arbitrary number of 
positive-helicity legs.

Using the CSW formulation,
we can write down an explicit expression for one such amplitude,
\def\cm#1{\kern -4pt#1\kern 1pt}
\def\pI{{\phantom1}}
\def\Ks#1{(-K_{#1})\vo}
\begin{eqnarray}
&& \hskip -.8 cm  A_n(1^-,2^-,3^+,4^-,5^+,\ldots,n^+) = \vphantom{\sum}
 \nonumber \\
&&{i\over s_{23}}  A_{3}(2^-,3^+,\Ks{2,3}^-)
   A_{n-1}(1^-,\Ks{1;4\cdots n}^+,4^-,\ldots,n^+)
\nonumber \\
&& +
\sum_{j_2=3}^{4}\sum_{j_1=j_2+2}^{n+1} 
    {i\over s_{j_2\cdots (j_1\!-\!1)}}
 A_{j_1-j_2+1}(j_2^+,\ldots,4^-,\ldots,(j_1-1)\vo^+,
                     \Ks{j_2\cdots (j_1\!-\!1)}^-)
\nonumber \\
&& \hphantom{+\sum_{j=m_2}^{m_3-1} {1\over s_{1\cdots j}}}
       \hskip 3mm\times
 A_{n-j_1+j_2+1}(1^-,2^-,\ldots,(j_2-1)\vo^+,
           \Ks{1\cdots (j_2\!-\!1);j_1\cdots n}_{\phantom1}^+\kern -4pt,
           j_1^+,\ldots,n^+)
\nonumber \\ 
&&+\sum_{j=5}^{n}
   {i\over s_{2\cdots (j-1)}}
   A_{j-1}(2^-,3^+,
         4^-,\ldots,(j-1)\vo^+,\Ks{2\cdots (j\!-\!1)}^+)
\label{ThreeMinusExample}
\\
&&\hphantom{+\sum_{j=m_2}^{m_3-1} {1\over s_{1\cdots j}}}\hskip 3mm\times
   A_{n-j+3}(1^-,\Ks{1;j\cdots n}^-,j^+,\ldots,n^+) \,,
\nonumber
\end{eqnarray}
where all indices are to be understood
$\mod n$.  
Each term corresponds to a different stripped diagram
in \fig{ThreeMinusFigure} with $(m_2,m_3)=(2,4)$,
and the sums on $j_{1,2}$ correspond to all possible ways of attaching 
the positive-helicity legs in between each pair of negative-helicity ones.

More generally, in the expression for a general amplitude with
three negative-helicity gluons, there are three double sums, each
term corresponding to a different stripped diagram,
\begin{eqnarray}
&A_n&(1^-,2^+,\ldots,
     m_2^-,(m_2+1)^+\cm,\ldots,m_3^-,
     (m_3+1)_\pI^+\cm,\ldots,n^+) = \vphantom{\sum}\nonumber\\
&&\sum_{j_1,j_2} 
           {i\over s_{j_2\cdots (j_1\!-\!1)}}
   A_{j_1-j_2+1}(\ldots,m_2^-,\ldots)
\,   A_{n-j_1+j_2+1}(\ldots,m_3^-,\ldots,1^-,\ldots)
\nonumber\\ &&+\sum_{j_1,j_2} 
    {i\over s_{j_2\cdots (j_1\!-\!1)}}
 A_{j_1-j_2+1}(\ldots,m_3^-,\ldots)
\, A_{n-j_1+j_2+1}(1^-,\ldots,m_2^-,\ldots)
\label{ThreeMinusGeneral}
\\ &&+\sum_{j_1,j_2}
   {i\over s_{j_2\cdots (j_1\!-\!1)}}
   A_{n-j_1+j_2+1}(1^-,\ldots)
   \,A_{j_1-j_2+1}(\ldots,m_2^-,\ldots,m_3^-,\ldots).
\nonumber
\end{eqnarray}
Each double sum over $j_1,j_2$ corresponds to the different ways of
attaching positive-helicity legs to a given stripped diagram.

\section{Skeleton Diagrams}
\label{SkeletonSection}

\begin{figure}[t]
\centerline{\epsfxsize 0.8 truein \epsfbox{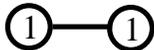}}
\caption[a]{\small 
The lone MHV skeleton diagram for amplitudes with three
negative-helicity gluons.
}
\label{ThreeMinusSkeletonFigure}
\end{figure}

We can go beyond stripping the diagrams of positive-helicity
legs to skinning them of negative-helicity ones too.  
This will yield what we shall call MHV {\it skeleton diagrams\/},
that is CSW diagrams
with all external legs removed.
We can obtain
the three stripped diagrams of \fig{ThreeMinusFigure} by starting
with the lone skeleton diagram shown in \fig{ThreeMinusSkeletonFigure}, 
and first attaching the negative-helicity
external gluons in all inequivalent ways consistent with the 
cyclic ordering, and the requirement that each vertex have exactly two
negative-helicity lines attached to it.  There are three independent ways
of doing this, each one corresponding to one stripped diagram.  We can
then attach all positive-helicity external gluons, again in all ways 
consistent with the cyclic ordering.  This gives rise to the double
sums in \eqn{ThreeMinusGeneral}.
Alternatively, we can attach all external legs in one step.
This leads to the following
representation for the same amplitude as in \eqn{ThreeMinusGeneral},
\begin{eqnarray}
&& A_n(1^+,\ldots,m_1^-,(m_1\!+\!1)\vo^+,\ldots,
m_2^-,(m_2\!+\!1)\vo^+,\ldots,
m_3^-,(m_3\!+\!1)\vo^+,\ldots,n^+) = \nonumber \\
&&\hskip 7mm
\sum_{j_1=1}^n\sum_{j_2=j_1+1}^{j_1-3}
{i\over s_{j_1\cdots j_2}} 
  A_{j_2-j_1+2\mod n}(j_1,\ldots,j_2,(-K_{j_1\cdots j_2})\vo^-)
\nonumber \\
&&\hskip 7mm\hphantom{ \sum_{j_1=1}^n\sum_{j_2=j_1+1}^{j_1-3}
                          {i\over s_{j_1\cdots j_2}} }\hskip 5mm\times
A_{j_1-j_2\mod n}(j_2+1,\ldots,j_1-1,(-K_{(j_2+1)\cdots(j_1-1)})\vo^+),
\label{CyclicThreeMinus}
\end{eqnarray}
where $n>4$; where only terms with one negative helicity between $j_1$
and $j_2$ (inclusively), and two negative helicities between $j_2+1$
and $j_1-1$ (inclusively), are included;
and where the inner sum and all subscripts
should be understood in a cyclic sense, for example
\begin{equation}
\sum_{j=n-4}^3 \equiv \sum_{j={(n-4)\cdots n,1\cdots3}}\qquad
{\rm\ and\ } \qquad \sum_{j=2}^{-2} \equiv \sum_{j=2}^{n-2}.
\label{CyclicSumDef}
\end{equation}

Both expressions~(\ref{ThreeMinusGeneral}) and~(\ref{CyclicThreeMinus})
 are of course equivalent to the
sum of all CSW diagrams.
Recall that to obtain a twistor-space
amplitude, one integrates both over a moduli space of curves in
$\CP^{3|4}$, and over the positions of the external
particles on the curves. As we shall see in the \sect{TwistorSection}, the
skeleton diagrams encode the type of curves in
$\CP^{3|4}$ over which one integrates.  The different ways of 
attaching the external
legs to the skeleton
correspond in twistor space to the arrangements of the external
particle insertions on the curves. The introduction of skeleton
diagrams is thus very natural from a twistor-space perspective.

\begin{figure}[t]
\centerline{\epsfxsize 3 truein \epsfbox{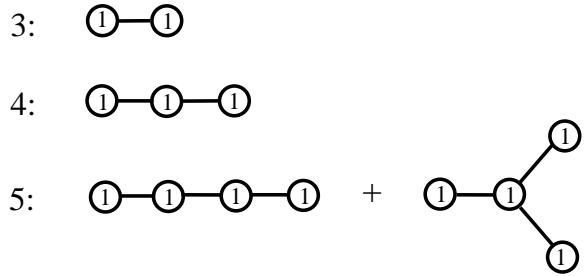}}
\caption{The skeleton diagrams with up to six external negative-helicity
legs.
The numbers inside the circles give the degree of the associated curves in 
twistor space.}
\label{SkeletonExamplesFigure}
\end{figure}

 In categorizing the skeleton
diagrams, we need to list only the topologically inequivalent ones.
 In \fig{SkeletonExamplesFigure} we list the distinct skeleton diagrams
for amplitudes with up to
five (external) negative helicities.  In general with $m$ negative
helicities, we must sum over all distinct topologies containing $d=m-1$
MHV vertices linked together in all possible trees.  As in
\eqn{CyclicThreeMinus}, we recover the sum over CSW diagrams from
the skeleton diagrams by summing over all ways of attaching external
legs --- positive and
negative helicities --- respecting the cyclic
ordering and the requirement that each MHV vertex have exactly two negative
helicity legs attached (including the internal ones).  Note that MHV vertices
with two negative-helicity legs attached but no positive-helicity ones
 vanish and should therefore not be included.

%

Of course, as the number of negative-helicity legs increases, the
number of topologically distinct skeleton diagrams grows exponentially.
Moreover, we will have  additional sums over ways of attaching legs
to vertices.  

Is there is a way to tame this growth?  In the context of
color-ordered Feynman rules, recurrence 
relations~\cite{Recurrence,LightConeRecurrence,DixonTASI}
 have proven a powerful
method of evaluating tree amplitudes, both for analytic and direct
numerical purposes.  A proper use of recurrence relations recycles
information and thereby reduces the
computational complexity of evaluating a color-ordered
amplitude (for given external helicities) from exponential to polynomial.
 One might imagine reformulating
the CSW rules in a recursive fashion, along the same lines as
the recasting of the color-ordered Feynman rules; but with the most
obvious approach 
there are vertices with an arbitrary number of legs.  This 
would lead to terms in the recurrence relations with an arbitrary number
of sums, which is what we are seeking to avoid.

There is another way to proceed, generalizing \eqn{CyclicThreeMinus},
which we find more in the spirit of the twistor-space structure of the
amplitudes and the CSW rules.  It also sheds light on the non-MHV
vertices that arose in the work of Gukov, Motl and
Neitzke~\cite{Gukov}.  First, we must give a more precise definition
of such vertices, which we do in the next section.


\section{Non-MHV Vertices}
\label{NonMHVVerticesSection}

The CSW construction of amplitudes uses MHV vertices, and is dual 
to integrating over
the moduli space of degree one curves in
$\CP^{3|4}$.  In contrast, Roiban, Spradlin,
and Volovich~\cite{RSV1,RSV2} compute amplitudes by integrating over 
the moduli space of 
one connected degree $d$ curve.  Gukov, Motl, and
Neitzke~\cite{Gukov} argued that the two approaches are
equivalent by showing that they can both be reduced to an integral over 
the common boundary of the two moduli spaces.
As a by-product of their analysis, they noted the existence of 
non-MHV vertices, though they did not give explicit 
expressions for them.  A non-MHV vertex corresponds to a twistor-space 
curve of degree $d = q -1$, where $q>2$ is the number of 
negative-helicity lines attached to the vertex.

Let us now make precise the notion of non-MHV vertices.
In the next section, we will show how to use them 
to reorganize the CSW construction in a recursive form.
Later, we shall also use
these vertices to show the equivalence of various twistor-space
prescriptions in a field-theory approach, and to determine any
associated combinatoric factors.

We can define an $n$-point degree-$d$ non-MHV vertex by
copying the set of CSW diagrams used to define an {\it on\/}-shell 
$n$-point amplitude with $d+1$ negative-helicity external legs,
and taking all external legs {\it off\/} shell.
 That is, a CSW
amplitude may be converted to a vertex,
\begin{equation}
A_n(1,2, \ldots, n) \longrightarrow V_n(1,2, \ldots, n),
\label{NonMHVVertexDef}
\end{equation}
by using the same off-shell prescriptions (\ref{CSWOffShell}) or
(\ref{OffShellPrescription}) on all external lines as used for sewing MHV
vertices together to form non-MHV amplitudes.  
It is important that the {\it same\/} light-cone reference
momentum used for defining off-shell
lines connecting the MHV vertices also
be used for the external legs; otherwise this reference
momentum will not necessarily drop out of on-shell amplitudes
we shall later construct
using the non-MHV vertices.  Moreover the prescription
(\ref{OffShellPrescription}) applies only to the MHV vertices,
and not to the propagators connecting them.  The latter
carry the momenta $K$ and not $k^\proj$ used to
define the spinors in the vertices.  The definition of
non-MHV vertices given here is not necessarily unique; in a sense, there
are at the very least different `gauge' choices corresponding to different
choices of the reference momentum $q$.

%
\begin{figure}[t]
\centerline{\epsfxsize 5.5 truein \epsfbox{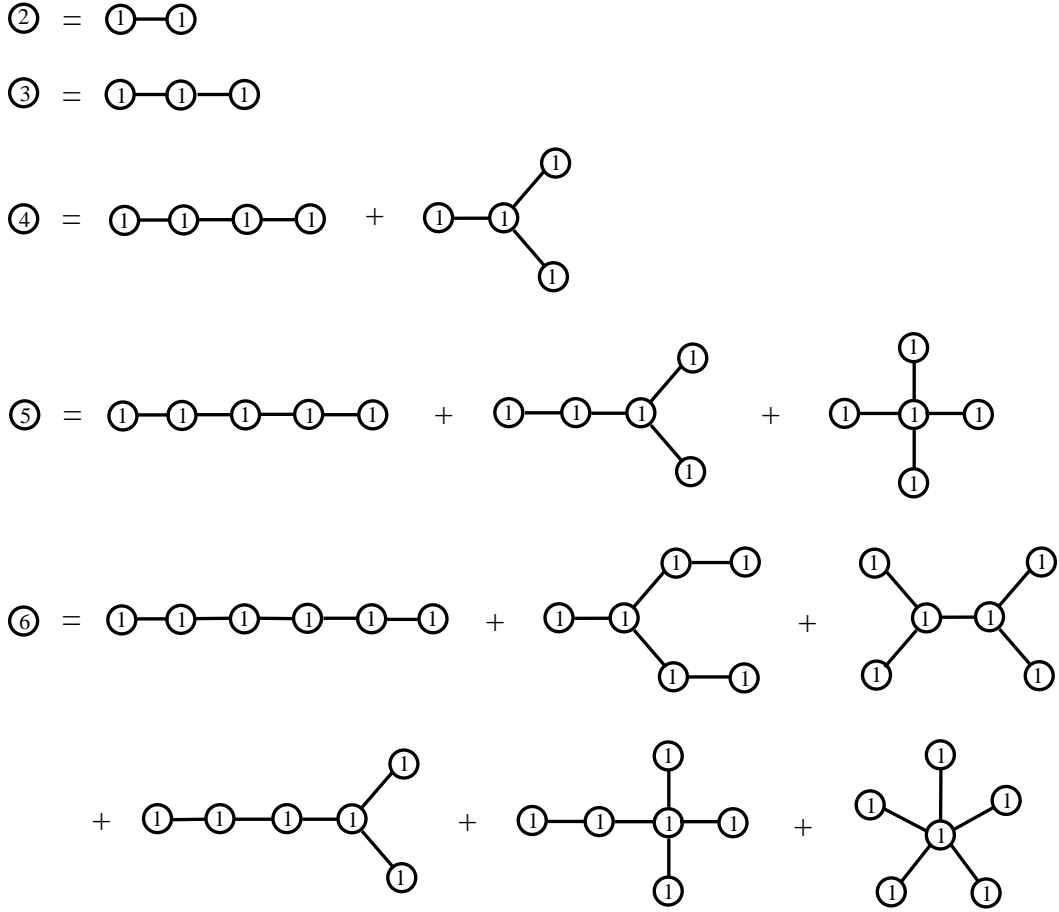}}
\caption{The skeleton diagrams for vertices up to degree six.
The degrees are indicated by the numbers inside the circles.
}
\label{VerticesFigure}
\end{figure}

We may represent the non-MHV vertices stripped of external lines in
terms of skeleton diagrams, as shown in \fig{VerticesFigure}.
Following the same discussion as for amplitudes, the full vertices are
obtained by dressing these skeleton diagrams with off-shell external
lines in all possible ways consistent with cyclic ordering, and such that
each degree-$d$ vertex has exactly $d+1$ negative helicity lines
attached.

Since they are obtained from the CSW diagrams, the non-MHV vertices
yield non-MHV amplitudes upon taking the on-shell limit for external momenta.
  With the original CSW off-shell continuation
(\ref{CSWOffShell}) one should of course
divide out the additional spinor factors, that is for an $n$-gluon
amplitude,
\begin{equation}
A_n(1, \ldots, n) = 
(-1)^n \prod_{p} \spb{q}.p^2 \, \prod_{m} {1\over \spb{q}.m^2 } \times
   V_n(1, \ldots, n) \Bigr|_{k_i^2 = 0} \,, 
   \hskip 3 cm   i = 1, \ldots, n, 
\label{OnShellCSW}
\end{equation}
where the products run over the positive-helicity legs $p$ and the
negative-helicity legs $m$.  The off-shell continuation
(\ref{OffShellPrescription}) is somewhat simpler since we can
recover on-shell
amplitudes from the vertices simply by taking the
on-shell limit of all external momenta,
\begin{equation}
A_n(1, \ldots, n) = 
   V_n(1, \ldots, n) \Bigr|_{k_i^2 = 0}\,,
   \hskip 3 cm   i = 1, \ldots, n.
\label{OnShellLimit}
\end{equation}
These equations are reflected in the representation of $A_n$ and $V_n$ by
the {\it same\/} skeleton diagrams.


\section{Recursive Formulation of Twistor-Inspired Rules}
\label{RecurrenceSection}

As discussed at the
end of section~\ref{SkeletonSection},
the most effective rearrangement of
CSW diagrams from a computational point of view 
is one that recycles as much information as possible as
one increases the number of negative-helicity legs.  For this we seek
a recursive approach.

In formulating such an approach, it is helpful to think 
of the CSW construction as summing over all inequivalent
multi-particle factorizations
of a gauge-theory amplitude.  In massless gauge theories, amplitudes
do not have (full) poles in two-particle invariants.  The limit 
$s_{ij}\rightarrow 0$ corresponds to a collinear limit, and 
amplitudes go as $1/\sqrt{s_{ij}}$ rather than $1/s_{ij}$ in this 
limit (the four-point amplitude is an exception, where
the limit corresponds to a forward-scattering singularity).  These
collinear singularities are captured by the MHV vertices.
The amplitudes do have singularities in $(n>2)$-particle invariants, which
are precisely the propagator poles prescribed by the CSW rules.
An expression for an amplitude, written in terms of MHV vertices and 
propagators, corresponds to the complete factorization in all simultaneously
factorizable multi-particle channels.  For non-exceptional configurations
of external momenta, the same reasoning applies to the non-MHV vertices
as defined in the previous section.

However, we could choose to factorize one channel at a time.  
This would decompose an amplitude into the product of
two simpler off-shell amplitudes
(or vertices), and one propagator.  Indeed, examining any skeleton
diagram, we could simply pick one of the internal lines.  The contributions
corresponding to that skeleton diagram (not yet a complete
amplitude) can of course be written as the
product of the propagator represented by the given internal line, and
the off-shell amplitudes (or vertices) on either side.  Similarly, we
can represent contributions to non-MHV vertices as products of lower-degree
vertices and a propagator joining them.

The seeming problem with such an approach is that there are many possible
choices of internal lines, and no clear reason to pick one over another.
The most natural
 solution is to sum over all of them.  This overcounts each skeleton
diagram by a factor of the number of internal lines; but this overcount
is the same for every skeleton diagram contributing to a given amplitude,
and hence can be divided out uniformly.  Restoring the external legs as
before,
this gives us a representation
of a vertex with $c$ negative helicities
in terms of two simpler off-shell vertices and a propagator,
\begin{eqnarray}
&& V_n(1^+,\ldots,m_1^-,(m_1\!+\!1)\vo^+,\ldots,
m_2^-,(m_2\!+\!1)\vo^+,\ldots,
m_c^-,(m_c\!+\!1)\vo^+,\ldots,n^+) = \nonumber \\
&&\hskip 7mm
{1\over (c-2)}\sum_{j_1=1}^n\sum_{j_2=j_1+1}^{j_1-3}
{i\over s_{j_1\cdots j_2}} V_{j_2-j_1+2 \mod n}
 (j_1,\ldots,j_2,(-K_{j_1\cdots j_2})\vo^{-}) 
\label{VertexRecursion}\\ 
&& \hskip 4mm \hphantom{ {1\over (c-2)}\sum_{j_1=1}^n\sum_{j_2=j_1+1}^{j_1-3}
                           {i\over s_{j_1\cdots j_2}} }\hskip 6mm\times
V_{j_1-j_2\mod n}(j_2+1,\ldots,j_1-1,(-K_{(j_2+1)\cdots(j_1-1)})\vo^+)\,,
\nonumber
\end{eqnarray}
where each term is included only if there is at least one 
negative-helicity gluon in the cyclic range $[j_1,j_2]$ and at least two
in the range $[j_2+1,j_1-1]$. (Again, as in \eqn{CyclicThreeMinus}, 
all indices are
to be understood $\mod n$, and all sums in a cyclic sense~\eqn{CyclicSumDef}.)
The two vertices are simpler in the sense that each has lower degree,
that is fewer 
negative-helicity legs (including the off-shell ones) than the parent
vertex.  This means that this equation provides a recurrence relation
for evaluating any tree-level non-MHV vertex in massless gauge theory,
and via \eqn{OnShellLimit}, for evaluating any tree-level
amplitude.  Note that the sum implicitly runs over different degrees
for the two vertices on the right-hand side, because the number of
negative-helicity external legs in $[j_1,j_2]$ can vary.

Using this recurrence identity, along with the MHV vertices of \eqn{Vertices}
with additional legs taken off shell, we can compute amplitudes numerically
as well as analytically.  It is straightforward to implement this equation
in computer code; the on-shell conditions expressed in \eqn{OnShellCSW}
or~(\ref{OnShellLimit}) can be imposed from the beginning of a computation.
Indeed, we verified numerically that \eqn{VertexRecursion} agrees
with a light-cone version of conventional recurrence relations through $n=9$.
Although we focus on gluon amplitudes here, with suitable sums over the
quantum numbers of internal lines, and fermion or scalar analogs of the
vertices of \eqn{Vertices}, the relation also applies to amplitudes
with other massless external states, in non-supersymmetric as well
as supersymmetric theories.

\begin{figure}
\centerline{\epsfxsize 4.5 truein \epsfbox{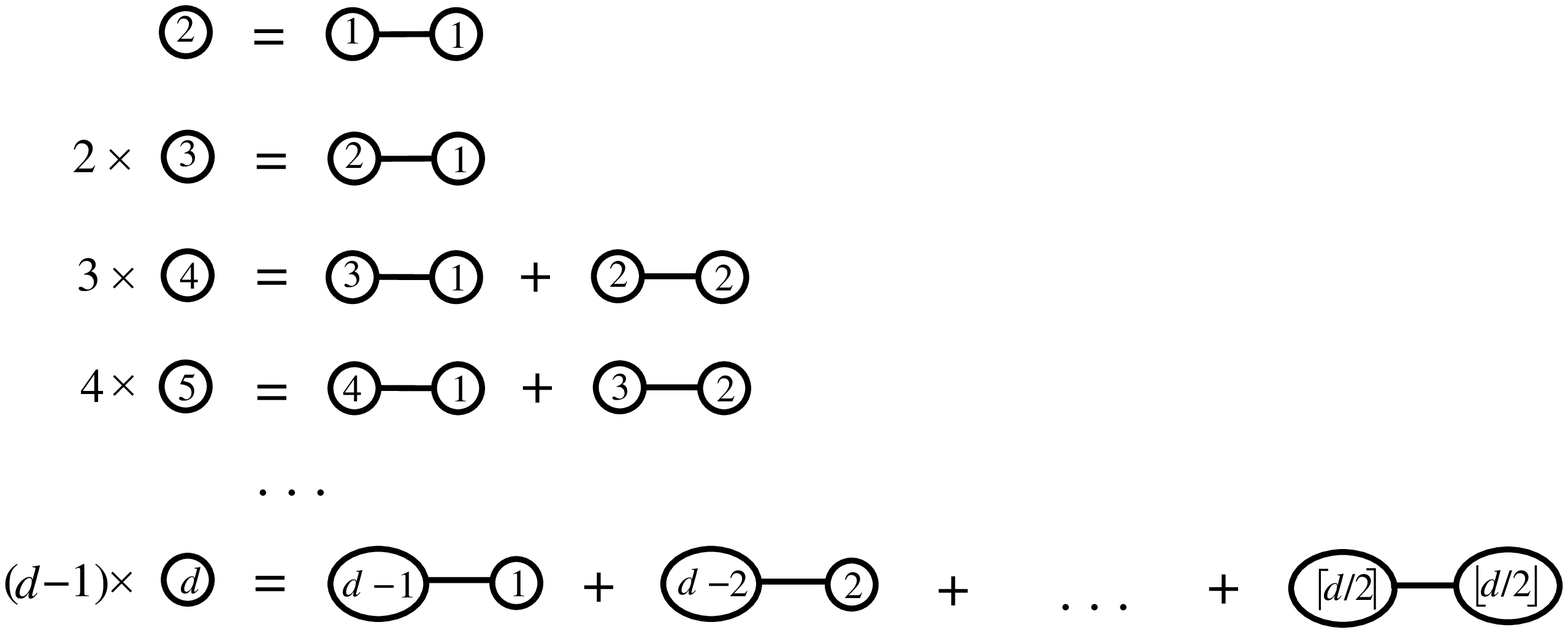}}
\caption{A recursive construction of non-MHV vertices.}
\label{RecursiveFigure}
\end{figure}

We illustrate \eqn{VertexRecursion} in \fig{RecursiveFigure}.
To follow the derivation pictorially, 
start with the complete MHV skeleton diagrams for non-MHV vertices
of various degrees, illustrated in \fig{VerticesFigure}.  Pick each 
propagator on the right-hand side in turn, and mark it.  Group and merge
the subdiagrams on either side of the marked propagator
into two non-MHV vertices with
(say) degrees $d_1$ and $d_2$ respectively, where $d_1+d_2 = d$, the
degree of the corresponding non-MHV vertex on the left-hand side.  This gives us 
one term on the right-hand side of the corresponding equation 
\ref{VertexRecursion}.  No matter what its topology, each skeleton
diagram contributing to a non-MHV vertex has $d-1$ internal propagators.
Hence summing over all possible ways of marking
propagators, and then grouping and merging subdiagrams,
overcounts by a {\it uniform\/} factor, which we must divide out
in \eqn{VertexRecursion} and \fig{RecursiveFigure}.  (Note
that each individual diagram in \fig{RecursiveFigure} actually represents
two different ways of attaching the negative-helicity external legs, because
there are two different helicity assignments possible for the internal 
line.)
We cannot simply pick one of the groupings, because we will not
obtain all required poles in a factorization, or equivalently because
some CSW diagrams would end up getting dropped.  Other ways
of decomposing non-MHV vertices will be discussed in \sect{AlternativeSection}.

%

As an explicit example, consider the degree-four
vertex.  In \fig{VerticesFigure}, we see that it is expressed
in terms of two MHV skeleton diagrams.  We isolate
propagators as above in two different ways, grouping subdiagrams into
either degree-three $\times$ degree-one diagrams
or into degree-two $\times$ degree-two
groups.  The first MHV
skeleton has two $3\times 1$ groupings while the second has three such
groupings.  In contrast,
there is only a single way to group the first MHV diagram into 
a $2 \times 2$ grouping, 
and no way to group the second in this manner.
Thus each MHV diagram is overcounted three times by the listed collection
of non-MHV skeleton diagrams: the first appears twice in 
$3\times 1$ subdiagrams, and once in $2\times 2$ subdiagrams, while the
latter shows up three times in $3\times 1$ subdiagrams.

Similarly, consider the degree-five vertex depicted in
\fig{VerticesFigure}.  This vertex 
may be expressed in terms of three MHV
skeleton diagrams.  In the recursive approach we group these into $4
\times 1$ + $3\times 2$ skeleton subdiagrams.  
We can decompose the first MHV skeleton diagram into $4\times 1$
subdiagrams in two ways; the second MHV skeleton diagram can be so
decomposed in three ways; and the third MHV skeleton diagram, in four ways.
The first skeleton diagram admits two decompositions into $3\times 2$
subdiagrams; the second only one such decomposition, and the last
skeleton diagram allows no such decomposition.  Summing over the
two different decompositions, we find a uniform overcount of a factor
of four.  More generally, for a degree-$d$ vertex, as
explained above, we obtain a uniform overcount of $d-1$ upon
summing 
over all groupings into pairs of non-MHV vertices.

The attentive reader may wonder how the combinatoric factors arise when
reversing this process, and expanding the non-MHV vertices back into MHV
vertices.  They arise in counting the number of ways that the diagram
can be re-expanded into MHV skeleton diagrams {\it after\/} reattaching
the external
negative-helicity gluons in order
to recover stripped diagrams.  The counting in this direction is thus
a bit more involved.

\section{Alternative Representations}
\label{AlternativeSection}

Besides the computationally useful recursive construction of the
previous section, other reorganizations of the MHV skeleton diagrams
are possible.  Each corresponds to a different combination of curves
in twistor space.
For example, one can organize the diagrams so
that a fixed number of vertices appear in each 
skeleton diagram; or
that only a given number of higher degree vertices appear; or in
other ways motivated by topological considerations.
The recursive construction presented in the previous section is an
example of the first kind of reorganization, with
the number of vertices fixed at two.  

\begin{figure}[t]
\centerline{\epsfxsize 3.5 truein \epsfbox{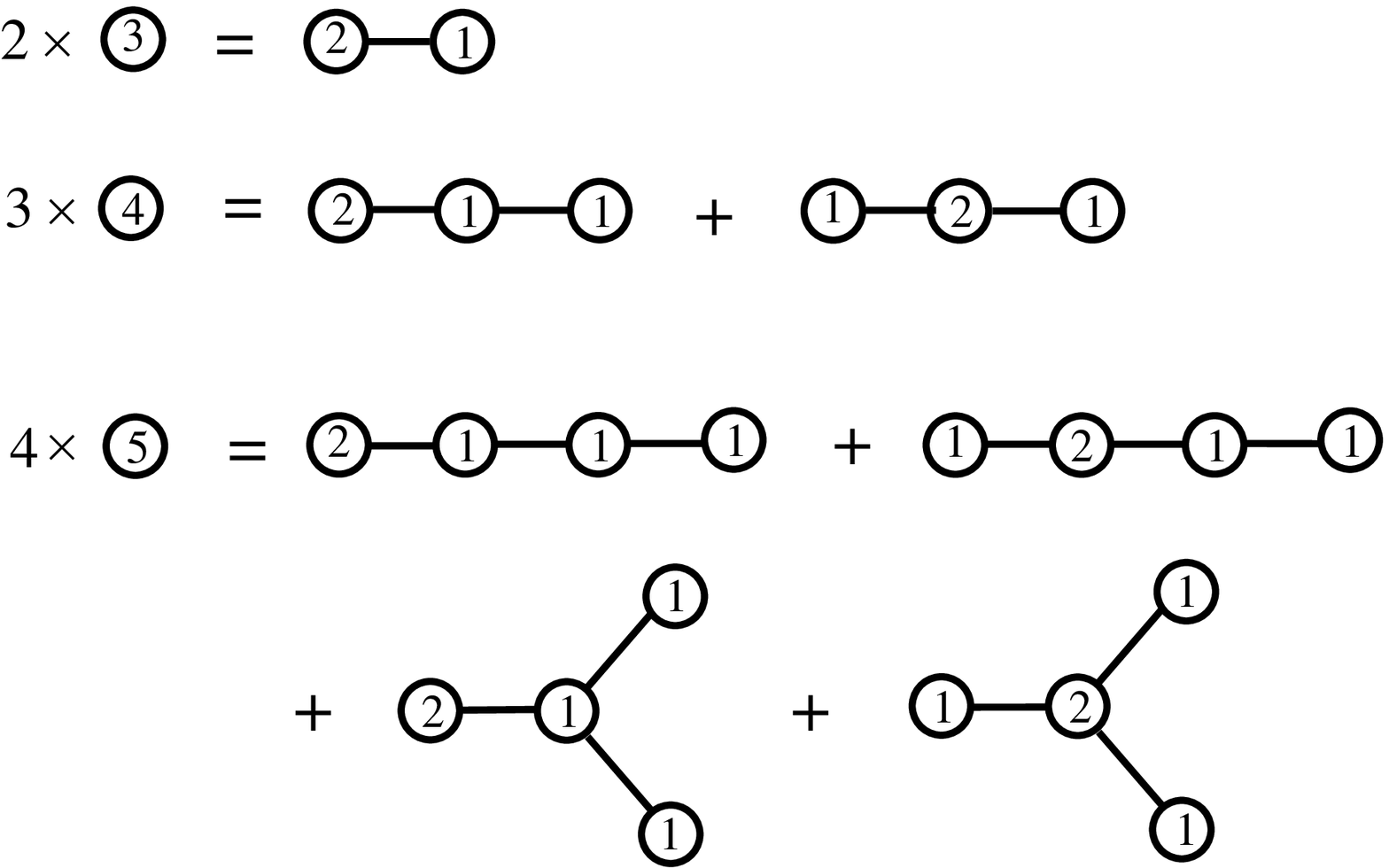}}
\caption[a]{\small The degree $d$ vertices can be expressed in terms
of one degree two vertex and $d-1$ degree one vertices. There is
a uniform overcount of $d-1$ in this case.}
\label{DegreeTwoExampleFigure}
\end{figure}

Each of these reorganizations may be obtained
by merging connected groups of MHV vertices into non-MHV vertices.
In each of these cases, we determine all combinatoric or
overcount factors directly.

As an example with fixed number of higher-degree vertices,
we can arrange to have a single degree-two vertex in each diagram.
This rearrangement may not be particularly useful, but does serve 
to illustrate
the principle.  This leads to the skeleton diagrams depicted in
\fig{DegreeTwoExampleFigure}.  These diagrams are obtained from the
ones in \fig{VerticesFigure} by merging a single pair of directly-connected
 MHV vertices in each diagram.  This gives us one degree-two vertex and
$d-2$ degree-one vertices as shown in \fig{DegreeTwoExampleFigure}.
If we sum over all possible pairings of linked MHV vertices, the
skeleton diagrams are again overcounted by a uniform factor of $d-1$ since
each diagram has $d-1$ linked pairs.  This overcount was already noted
in ref.~\cite{Gukov}.  Similarly, if we create non-MHV diagrams for
a degree-$d$ vertex by
grouping and merging MHV vertices so as to collapse $r$ links
in all possible ways, the
overcount factor is ${d-1\choose r}$.

%
\begin{figure}[b]
\hskip -1truecm
\centerline{\epsfxsize 5 truein \epsfbox{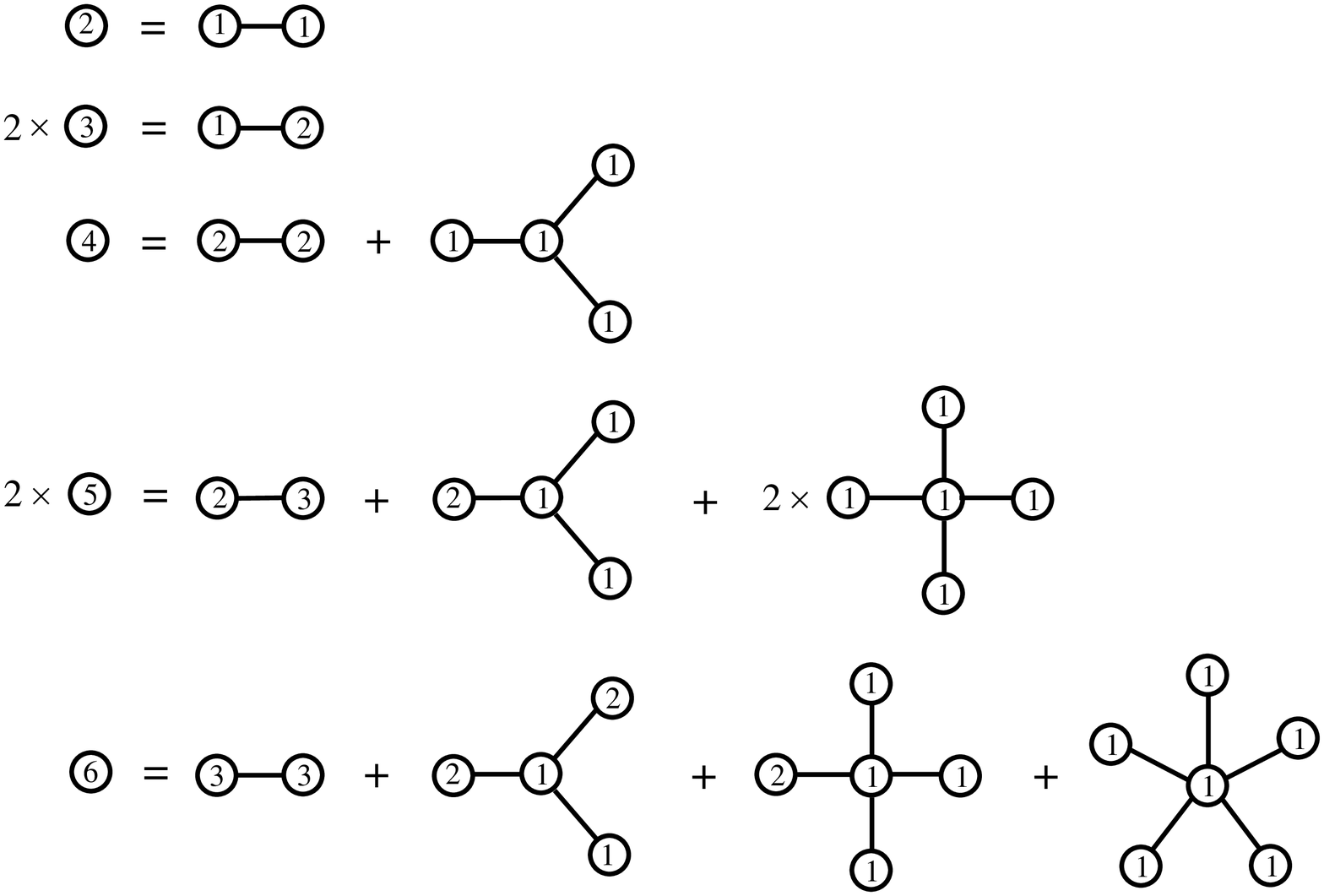}}
\caption{An alternative set of skeleton diagrams for obtaining non-MHV 
vertices. In this case, for even degree there is no overcount,
while for odd degree there is an overcount by a factor of two. 
In addition there are relative combinatoric factors between the skeleton
diagrams.
}
\label{AlternativeFigure}
\end{figure}

A more subtle example arises from trying to balance the degrees
of the vertices in a diagram.  
This reorganization is illustrated in \fig{AlternativeFigure}.
From a twistor-space point of view, it
expresses an amplitude as a sum of contributions with differing
number of curves.  One contribution to
the degree $d$ vertex has a propagator connecting 
a degree-$\lfloor d/2\rfloor$ to a degree-$\lceil d/2\rceil$ vertex,
along with a series of `bicycle wheel'
diagrams with vertices up to degree
$\lfloor (d-1)/2\rfloor$ connected to a central vertex of degree one.
($\lfloor x\rfloor$ is the standard notation for the largest integer
less than or equal to $x$, and $\lceil x\rceil$ is that for the smallest 
integer larger than or equal to $x$.)

This example also illustrates the possibility of relative
combinatoric factors.  For $d$ even there are no combinatoric or overcount
factors while for $d$ odd there is an overall factor of two, as well as
relative factors of two between different diagrams.

To derive the combinatoric factors within any given
reorganization, we must count the total
number of ways each MHV skeleton diagram can be merged to give
any of the skeleton diagrams built using non-MHV vertices.  We may need to add
additional copies of some of the latter diagrams in order to obtain
a uniform overcount factor, which we then divide out.

For the degree-five vertex, for example, there are two ways to group
and merge the linear skeleton appearing first on the 
fourth line of
\fig{VerticesFigure} into the connected product of a degree-two and a 
degree-three vertex shown first on the penultimate line of
\fig{AlternativeFigure}.  There is one way to group and merge the middle
`lobster' diagram into such a product, and one way to group and merge
it into the middle diagram on the penultimate line of this figure.
There is no grouping to be performed on the last skeleton appearing
in \fig{VerticesFigure}, so it simply appears as is in \fig{AlternativeFigure}.
However, in order to obtain a uniform overall
combinatoric factor, we must take
two identical copies of it.  To obtain the five vertex,
the sum of these non-MHV skeleton diagrams must of course be divided by
the overall combinatoric factor of two.  As illustrated by the degree-six
vertex, this patterns continues in the general form mentioned earlier.

To recover full non-MHV diagrams, we must restore the external legs.
Conceptually, we can think of doing this in two steps: first, obtain
a set of stripped diagrams by attaching
negative-helicity external legs to the skeleton in all possible ways consistent
with the cyclic ordering and the degree of the vertices (a degree-$d$
vertex must have $d+1$ negative-helicity gluons including the internal
ones).  One then attaches the positive-helicity external legs in all
possible ways consistent with the cyclic symmetry.  Alternatively, we
can simply attach the external legs in all ways consistent with cyclic
symmetry and the degree of the vertices.

\def\Nm{N_{-}}
Doing so for the decomposition of
the degree-four vertex on the third line of \fig{AlternativeFigure}, 
for example, we obtain the following expression,
\begin{eqnarray}
&& V_n^{\hbox{\textcircled{\scriptsize 4}}}%
(1^+,\ldots,m_1^-,(m_1\!+\!1)\vo^+,\ldots,
m_2^-,(m_2\!+\!1)\vo^+,\ldots,
m_5^-,(m_5\!+\!1)\vo^+,\ldots,n^+) = \nonumber \\
&&\hskip 5mm\sum_{j_1=1}^n
\sum_{j_2=j_1+1\atop \Nm{[j_1,j_2]}=2}^{j_1-3} {i\over s_{j_1\cdots j_2}}
V_{j_2-j_1+2 \mod n}
 (j_1,\ldots,j_2,(-K_{j_1\cdots j_2})\vo^{-}) 
\nonumber\\ 
&& \hskip 5mm \hphantom{ \sum_{j_1=1}^n
\sum_{j_2=j_1+1\atop \Nm{[j_1,j_2]}=2}^{j_1-3} {1\over s_{j_1\cdots j_2}}%
      }\hskip 6mm\times
V_{j_1-j_2\mod n}(j_2+1,\ldots,j_1-1,(-K_{(j_2+1)\cdots(j_1-1)})\vo^+)
\nonumber\\
&&-
\sum_{j_1=1}^n
\sum_{j_2=j_1+1\atop \Nm{[j_1,j_2]}=1}^{j_1-5}
\sum_{j_3=j_2+1}^{j_1-4}
\sum_{j_4=j_3+1\atop 1\leq\Nm{[j_3,j_4]}\leq 2}^{j_1-3}
\sum_{j_5=j_4+1}^{j_1-2}
\sum_{j_6=j_5+1\atop 1\leq\Nm{[j_5,j_6]}\leq 2}^{j_1-1}
{i\, c(\Nm{[j_3,j_4]},\Nm{[j_5,j_6]})
 \over s_{j_1\cdots j_2}s_{j_3\cdots j_4}s_{j_5\cdots j_6}} 
\label{AlternativeVertex}\\ &&\hskip 20mm\times
V_{j_2-j_1+2 \mod n}
 (j_1,\ldots,j_2,(-K_{j_1\cdots j_2})\vo^{-}) 
\,V_{j_4-j_3+2 \mod n}
 (j_3,\ldots,j_4,(-K_{j_3\cdots j_4})\vo^{\pm}) 
\nonumber\\ &&\hskip 20mm\times
V_{j_6-j_5+2 \mod n}
 (j_5,\ldots,j_6,(-K_{j_5\cdots j_6})\vo^{\pm}) 
\nonumber\\ &&\hskip 20mm\times
V_{j_1+j_3+j_5-j_2-j_4-j_6\mod n}
  (K_{j_1\cdots j_2}\vo^{+},(j_2+1),\ldots,(j_3-1),K_{j_3\cdots j_4}\vo^{\mp},
     (j_4+1),\ldots,
\nonumber\\ &&\hskip 20mm \hphantom{ \times
V_{j_1+j_3+j_5-j_2-j_4-j_6\mod n}
  ()}
(j_5-1),K_{j_5\cdots j_6}\vo^{\mp},(j_6+1),\ldots,(j_1-1))
\nonumber%
\end{eqnarray}
where as in \eqn{CyclicThreeMinus}, all indices should be understood $\mod n$,
and sums in a cyclic sense~\eqn{CyclicSumDef} (note, however, that in
the last vertex, $j_2+1,\ldots,j_3-1$ indicates the legs in between
$j_2$ and $j_3$, that is {\it no\/} legs if $j_3-j_2=1\mod n$).  Also,
the notation $\Nm{[a,b]}$ denotes the
number of negative-helicity gluons 
in the cyclic range $[a,b]$
(the notation $\Nm{[a,b]}=d$ in the summation indicates that one should
sum only over indices satisfying the constraint), and $c$ is a combinatoric
factor,
\begin{equation}
c(n_1,n_2) = {1\over 5-n_1-n_2}.
\end{equation}
This combinatoric factor arises from our choice of combining the
topologically distinct stripped diagrams into a single set of sums, and
has nothing to do with combinatoric factors arising from the reorganization
of MHV skeleton diagrams into non-MHV ones.  It would arise in the CSW
construction as well if we made a similar choice of combining the sums.
Indeed, the sixfold sum in the second term is exactly the same as would
emerge in a direct CSW computation.
Like the recurrence relation~(\ref{VertexRecursion}), 
\eqn{AlternativeVertex} could be used directly
for numerical computations, although we would expect \eqn{VertexRecursion}
to be more efficient for practical applications.

It is clear that many possible rearrangements of MHV vertices into
higher degree ones are possible.  These alternative representations
are useful for shedding light on the variety of instantonic
contributions of the topological string that properly reproduce the
amplitudes.  Any reorganization of
into non-MHV vertices is allowed, so long as one arranges for
overcount of CSW diagrams to be uniform.  The skeleton diagrams provide a
simple means for determining the overcounts and any relative
combinatoric factors that may be needed.


\section{Relation to Degenerating Curves in Twistor Space}
\label{TwistorSection}

As we have seen in the previous sections, it is easy to construct
intermediate prescriptions by merging selected groups of MHV vertices 
into non-MHV vertices. In this
section we relate this field-theoretic
construction of intermediate prescriptions to the twistor space
construction of Gukov et.\ al.~\cite{Gukov}.

In section \ref{SkeletonSection} we obtained skeleton diagrams by
skinning the MHV diagrams of all their external legs. As mentioned
there, this corresponds in twistor space to focusing on the types of
instantons one integrates over, and ignoring the external particle
insertions.

The duality between skeleton diagrams and curves in twistor space is
straightforward.  A skeleton diagram containing only degree-one
vertices reproduces a collection of CSW diagrams after reattaching the
external legs, and is dual to integrating over the moduli
space of intersecting degree-one curves in $\CP^{3|4}$. 
A vertex of degree $d$ in
a skeleton diagram corresponds to a twistor space curve of degree
$d$. When two vertices are connected by a line, the corresponding
twistor space curves intersect. Vertices which are not directly connected
correspond to non-intersecting curves.

The ways in which an MHV skeleton diagram  can give rise to
one of the non-MHV skeleton diagrams in the desired reorganization are
in one-to-one correspondence to the ways a collection of
intersecting lines can arise (via degeneration) in the integral
over higher-degree curves in
twistor space. Skeleton
diagrams make counting and indexing the way a curve degenerates 
straightforward.
 This allows writing a variety of intermediate prescriptions,
as exemplified in \sect{AlternativeSection}, keeping track of
combinatoric factors.

\begin{figure}[t]
\centerline{\epsfxsize 4.5 truein \epsfbox{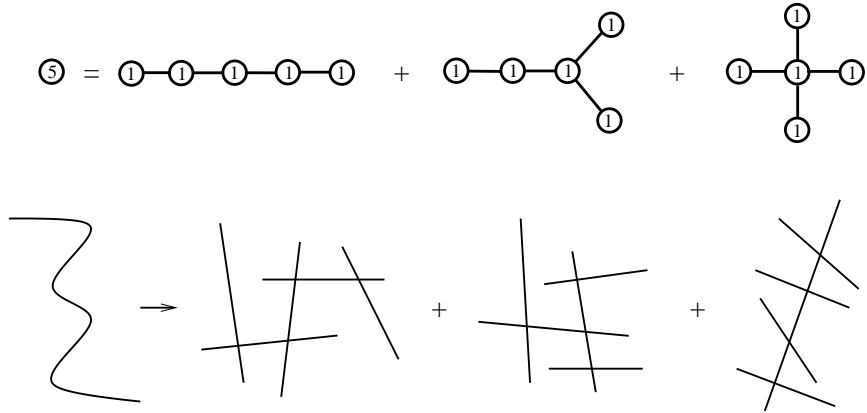}}
\caption{The MHV skeleton diagrams for amplitudes 
with six negative-helicity gluons, 
and the corresponding degenerations of a twistor space curve of degree five.}
\label{mhv-int2}
\end{figure}

As an example, let us focus on the degree-five vertex, which gives rise
to amplitudes with six negative-helicity legs.
 As we have seen, there are quite a few ways to
obtain this vertex starting from lower-degree vertices. One can use 
five MHV vertices as in Section \ref{SkeletonSection}, or sum over
combinations of two vertices of degrees $d_1+d_2= 5$ as in Section
\ref{RecurrenceSection}, or over other combinations of vertices as
explained in Section \ref{AlternativeSection}.

In twistor space, one can similarly obtain this amplitude by
integrating over the moduli space of connected degree-five instantons;
over the moduli space
of all possible pairs of
instantons of degrees $d_1+d_2=5$ linked by a twistor space
propagator; or over the moduli space of
five degree-one instantons linked by four propagators,
etc. As argued in ref.~\cite{Gukov}, the fact that all these integrals give
the same result comes from the fact that they all reduce to integrals
over a common locus. Roughly speaking, the integral over the moduli
space of connected degree-five instantons reduces to one over the part
of the moduli space where these curves degenerate into five
intersecting lines, while the integral over five lines connected by
propagators picks up a contribution from where the propagators shrink
to zero size. The integrals appearing in other prescriptions also
reduce to integrals over the locus of intersecting degree one curves.

\begin{figure}[t]
\centerline{\epsfxsize 3.5 truein \epsfbox{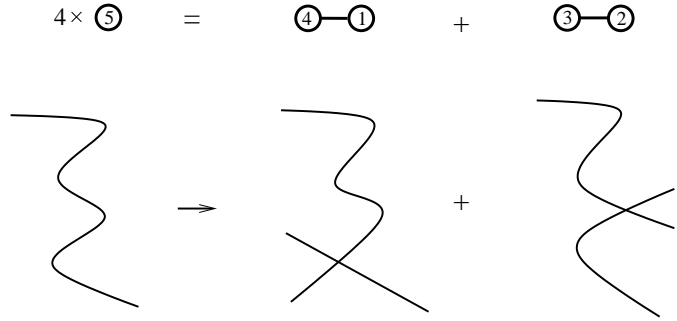}}
\caption{Skeleton diagrams for the recursive formulation of amplitudes 
with six negative-helicity legs, and
the corresponding degenerations of the degree-five instanton.}
\label{mhv-int1}
\end{figure}

\begin{figure}[b]
\centerline{\epsfxsize 4.4 truein \epsfbox{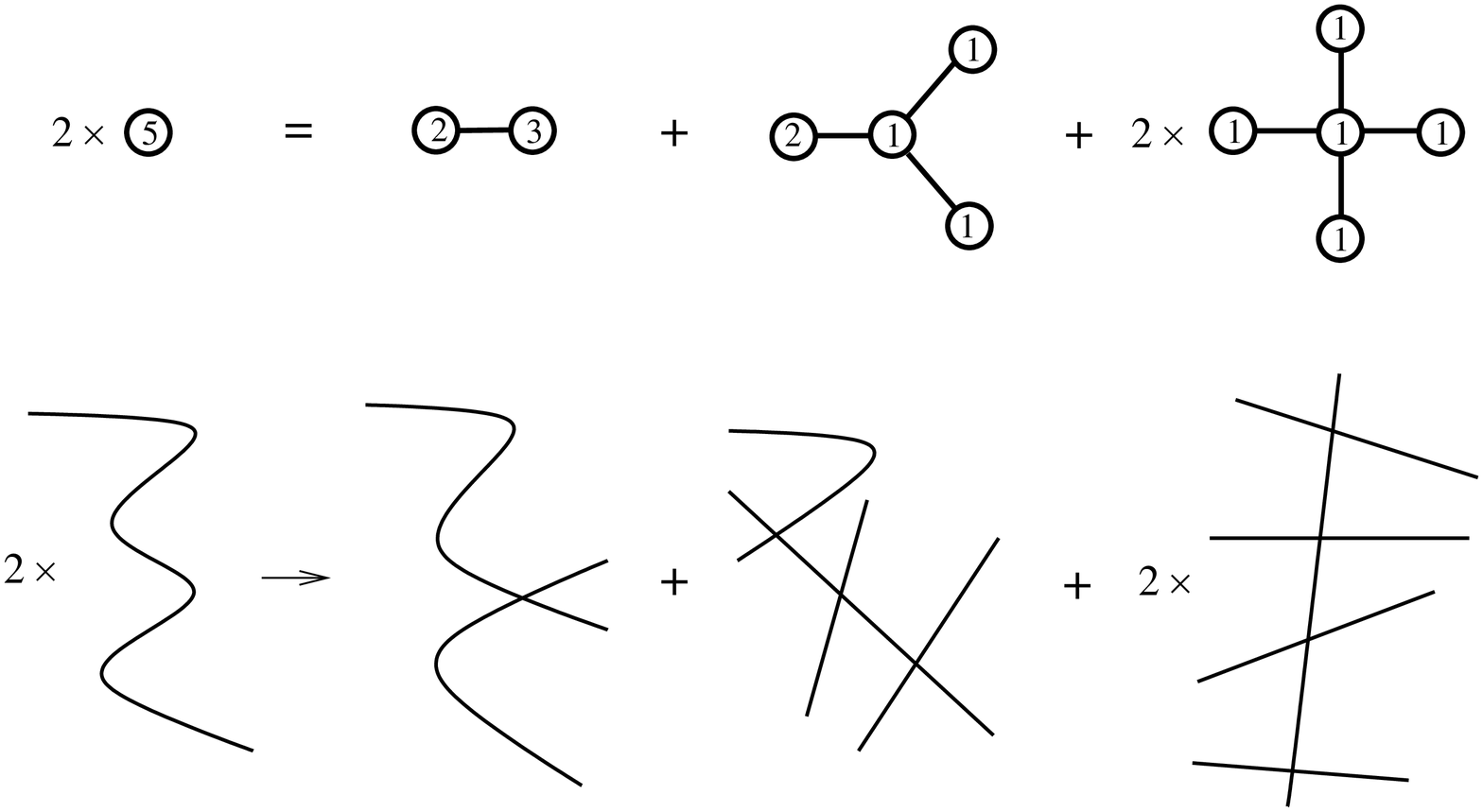}}
\caption{An alternative prescription for amplitudes with six 
negative-helicity legs.}
\label{mhv-int3}
\end{figure}

Thus, the intermediate prescriptions we formulate using skeleton
diagrams translate directly to twistor space prescriptions in which
one integrates over any combination of curves with the right degree,
such that the configurations of intersecting lines these curves
degenerate into are counted with the same weight. 
For example, In
\fig{mhv-int2} we show the degenerations of degree-five
curves into degree-one curves, along
with the corresponding skeleton diagrams --- which in this case
are nothing but CSW diagrams.  The intermediate prescription in
\fig{mhv-int1} consists of summing over two-instanton
configurations, with equal weight, and is dual to the recursive
formulation discussed in \sect{RecurrenceSection}. One can also
use more unusual representations of non-MHV vertices in
\sect{AlternativeSection} to engineer intermediate
prescriptions in which different numbers of instantons appear, as in
\fig{mhv-int3}.

In the recursive formulation, in
order to obtain amplitudes with $d+1$ negative-helicity legs,
one needs to sum over all combinations of two instantons with degrees
$d_1$ and $d - d_1$. Since this sum gives a large overcount, one might
naively expect that a single term in this sum gives the correct
result. However, for $d\geq 4$ this is never the case.  This
can either be shown by using skeleton diagrams, or by examining
twistor-space instantons.

To see this explicitly, consider the amplitude with six negative-helicity
legs discussed
above, and imagine trying to obtain it from integrating only over
two-instanton configurations of degrees three and two, ignoring
two-instanton configurations of degrees one and four.  As one can
see from \fig{mhv-int1}, the integral over two instantons of
degree three and two misses contributions coming from the last term of
the right-hand side.  This happens because two intersecting curves of
degrees two and three can never degenerate into a configuration of lines
where one line intersects all the other four. This configuration can
only be obtained by degenerating two intersecting curves of degrees one 
and four.

As discussed in \Sect{RecurrenceSection}, although this 
configuration does not appear in the
2--3 degeneration, it does appears in the 1--4 degeneration 
with a bigger overcount than the diagrams coming from the 2--3 degeneration,
so that after summing over all degenerations as in \fig{mhv-int1}
one obtains the full result with a uniform overcounting factor.

\begin{figure}[t]
\centerline{\epsfxsize 4. truein \epsfbox{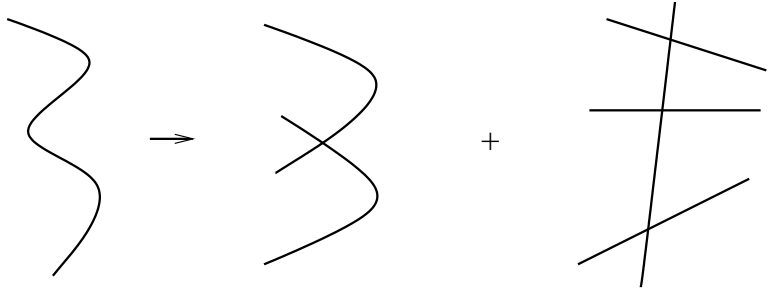}}
\caption{The degenerations of a twistor space curve of degree four 
corresponding to the intermediate prescription in \eqn{AlternativeVertex}.}
\label{deg4}
\end{figure}

Hence, the skeleton diagrams introduced earlier capture all information 
about the
ways in which moduli space curves degenerate, making it easy
to count these degenerations, and to produce novel intermediate 
prescriptions. The twistor-space degenerations corresponding to the
alternative decomposition of \eqn{AlternativeVertex}, for example,
are displayed in \fig{deg4}.
One can also use these diagrams (as in the example displayed in
\fig{mhv-int3}) to indirectly predict the ways in which complicated 
twistor-space
curves must degenerate.


\section{Conclusions}

In this paper, we have presented explicit formul\ae{} for
non-MHV vertices in the twistor-motivated formulation of gauge theory
amplitudes.  We have constructed them using skeleton diagrams, which
are CSW diagrams with external legs removed.  We have also made use of
these vertices to directly formulate gauge theory 
`intermediate' prescriptions~\cite{Gukov}
for computing
tree-level amplitudes.  Most notable amongst these intermediate
prescriptions are those with only two vertices in any diagram, which
give a recursive expression for degree-$d$ vertices in terms of lower-degree
ones.  Recursive methods have proven to be powerful tools in more
conventional diagrammatic approaches to gauge theories, and the recursive
formulation given here may well be the most computationally
efficient way to reorganize CSW diagrams.  The straightforward relation
between non-MHV vertices and amplitudes means that the recursive
approach offers a powerful method of computing amplitudes.
In particular, it allows the computation of amplitudes with an
arbitrary number of negative-helicity gluons without the need to keep
track of increasingly complicated graph topologies that would arise
in MHV diagrams.  It would be interesting to solve the recurrence relation
explicitly.

Skeleton diagrams correspond to collections of twistor-space curves.
Different skeleton diagrams, built out of different
vertices, for the same non-MHV amplitude correspond to different intermediate
prescriptions for the curves over which one integrates
in order to compute that amplitude in twistor space.  The
descriptions of amplitudes via skeleton diagrams and via twistor-space
curves can be thought of as dual.  Indeed,
our approach, proceeding from CSW diagrams, can be thought of as
complementary to the approach of Gukov et al.~\cite{Gukov}.  From
CSW diagrams, we proceed by defining
explicit forms of the non-MHV vertices, on to an understanding of how 
the integrals over moduli spaces of curves should reduce to the consideration of
degenerate curves.  This approach allows us to determine combinatoric
factors explicitly.
It also makes the construction of intermediate
prescriptions less mysterious --- ultimately, all the intermediate
prescriptions are just clever ways of grouping CSW diagrams and merging
subdiagrams into vertices.  The analysis here illustrates a practical
duality between field-theory and topological string-theory approaches.  
Certain aspects of amplitudes, such as the existence of a CSW construction,
are much clearer in twistor space, whereas details of the
construction of non-MHV vertices are clearer in field theory.

Although tree-level amplitudes correspond only to the classical theory,
deeper understanding of these amplitudes has in the past powered new
calculations of loop amplitudes.  These explicit calculations have revealed 
unexpected simplicity and intriguing relations that could not have been
guess on general field-theoretic grounds.  We can look forward to new
results and structures in the quantum theory based on calculations emerging
from the CSW construction and related work.

\section*{Acknowledgments}

We thank the KITP at Santa Barbara, where this
work was initiated, for its hospitality. 
We would also like to thank Sergio Ferrara, Radu Roiban, and Peter Svrcek 
for helpful discussions. 

This research was supported in part by the US Department of Energy under
contract DE--FG03--91ER40662, and in
part by the National Science Foundation under grants PHY99--07949, 
PHY00--99590 and PHY01--40151.
Any opinions, findings 
and conclusions expressed in this article
are those of the authors and do not necessarily reflect the views
of the National Science Foundation or other government agencies.


\end{document}